\documentclass[12pt]{article}
\usepackage{url} 
\usepackage{soul}

\usepackage{amsmath}
\usepackage{amssymb}
\usepackage{graphicx}
\usepackage{hyperref}
\usepackage{tikz}
\usetikzlibrary{positioning}
\usepackage{url}
\usepackage{natbib}
\usetikzlibrary{shapes.geometric}
\usepackage{multirow}
\usepackage{siunitx}
\usepackage{subcaption} 
\usepackage{tikz}
\usepackage{booktabs}
\usepackage[export]{adjustbox}

 \usepackage{tabularx}
 \usepackage{array}

\usepackage[normalem]{ulem}

\usepackage{appendix}
\usepackage{authblk} 
\newenvironment{keypoints}{%
  \vspace{10pt}
  \noindent\textbf{Key Points:}
  \begin{itemize}
    \setlength{\itemsep}{2pt}
    \setlength{\parskip}{0pt}
}{%
  \end{itemize}
  \vspace{10pt}
}

\usepackage[margin=0.5in]{geometry} 

\begin{document}

%
%


%
%




\title{Defects and Inconsistencies in Solar Flare Data Sources: Implications for Machine Learning Forecasting}
\author[1]{Ke Hu}
\author[1]{Kevin Jin}
\author[1]{Victor Verma}
\author[2]{Weihao Liu}
\author[2]{Ward Manchester IV}
\author[2]{Lulu Zhao}
\author[2]{Tamas Gombosi}
\author[1]{Yang Chen\thanks{Corresponding author: ychenang@umich.edu}}

\affil[1]{Department of Statistics, University of Michigan, Ann Arbor}
\affil[2]{Department of Climate and Space Sciences and Engineering, University of Michigan, Ann Arbor}







\maketitle

\begin{keypoints}
\item Widely used flare catalogs exhibit substantial discrepancies in event occurrence, reported intensities, and active-region association. 
\item These inconsistencies can affect the evaluated performance of data-driven flare-forecasting models.
\item We offer concrete recommendations regarding the selection of data products for operational flare forecasting.
\end{keypoints}

%
%

%
%


\begin{abstract}
Machine learning models for forecasting solar flares have been trained and evaluated using a variety of data sources, including Space Weather Prediction Center (SWPC) operational and science-quality data. Typically, data from these sources is minimally processed before being used to train and validate a forecasting model. However, predictive performance can be affected if defects and inconsistencies between these data sources are ignored. For a number of commonly used data sources, along with the software that queries and outputs processed data, we identify their defects and inconsistencies, quantify their extent, and show how they can affect predictions from data-driven machine-learning forecasting models. We also outline procedures for fixing these issues or at least mitigating their impacts. Finally, based on thorough comparisons of the trained forecasting model's predictive skill scores {with different data sources}, we offer recommendations for using different data products in operational forecasting. 
\end{abstract}

\section*{Plain Language Summary}
This paper shows that commonly used solar-flare datasets do not always agree, and that these disagreements can significantly distort how well machine-learning flare forecasts perform. We carefully compare and reconcile multiple flare catalogs, resolve known calibration issues, and build a cleaner reference dataset with improved flare locations and active region labels. When standard ML forecasting models are applied to the improved data, we gain practical guidance on which data products operational forecasting systems should use.

%
%

%


%
%
%
%

\section{Introduction}


Machine learning methods have been applied with great success to a variety of space weather forecasting problems, such as solar eruption forecasting~\citep[e.g.][]{Leka:2018b,leka2019comparison,chen2019iden,kasapis2022interpretable,whitman2023review,chen2025decent}, geomagnetic index prediction~\citep[e.g.][]{iong2022new,hu2023multi,nair2023magnet,iong2024sparse,chen2025geodgp}, and the prediction of terrestrial impacts of space weather events~\citep[e.g.][]{wang2019machine,telloni2023prediction,wang2023forecast,elfiky2025exploring,sun2025matrix}; see \citet{camporeale2018machine} and \citet{Camporeale:2019a} for reviews of methods and applications. In particular, machine learning methods have been used for solar flare forecasting, showing superior performance as compared with state-of-the-art physics-based models~\citep[e.g.][]{florios2018forecasting,liu2019predicting,sun2021improved,georgoulis2021flare,nishizuka2021operational,sun2022predicting,zheng2023multiclass,pandey2023unveiling}. 

A machine learning forecasting method typically predicts the value of a response variable from one or more predictor variables. In solar flare forecasting, in particular, the response is typically the flare class~\citep{nishizuka2018deep} or maximum soft X-ray flux during a flare~\citep{yi2020forecast}; the predictors may include remote sensing images such as those from the Solar Dynamics Observatory (SDO)~\citep[][]{Boerner:2012a,lemen2012atmospheric,pesnell2012,woods2012,sun2023tensor} or extracted features like the Space eather HMI Active Region Patch (SHARP) parameters~\citep[][]{SHARPs}. The method is trained to generate predictions by minimizing a loss function that quantifies the extent to which predicted outcomes deviate from observed outcomes. During training, the mapping from predictors to predictions is optimized (e.g., via gradient-based methods) to minimize the loss. Common choices of loss function in the flare forecasting literature include the cross-entropy loss (for binary classification of flares as strong or weak, e.g., \citet{chen2019iden}), sum of squares loss (for prediction of the logarithm of the peak flux, e.g., \citet{jiao2020solar}), and the negative log-likelihood (in either classification or regression tasks with statistical models, e.g., \citet{sun2023tensor}). State-of-the-art binary classification methods for forecasting flares include discriminant analysis~\citep{leka2018nwra}, logistic regression~\citep{korsos2021testing}, support vector machines~\citep{bobra2015sola}, random forests~\citep{liu2017predicting}, gradient-boosting machines~\citep{cinto2020solar}, long short-term memory (LSTM) neural networks~\citep{liu2019predicting}, and other more advanced network architectures~\citep[e.g.][]{zheng2019solar,sun2022predicting,Pandey2024}. 

The literature shows that the performance of machine learning methods is more constrained by the train, validation, and test design (random split versus chronological split, for example) and sample size (collecting non-overlapping flare events versus rolling window definition of maximum flaring events) than by the particular machine learning method adopted, due to solar flare data sparsity, especially strong M/X class flares; see \citet{Wang2020lstm,sun2022predicting,chen2024solar} for more discussions. Therefore, in this paper, instead of exhaustively evaluating all available machine learning methods, we select representative ones (to be detailed later) and focus on using them to calibrate the impact of the data sources (from different databases) and quality (scientific or operational) we use for training the machine learning model.

\newpage
\begin{figure}[!h]
    \centering
    \input{schematic}
    \caption{A schematic illustrating how a machine learning method produces predictions from predictors. Predictors are commonly computed from either images of active regions or summary statistical parameters on those regions. The upper-left part of the schematic depicts predictors computed from HMI and AIA images, represented as tensors with dimensions $h \times w \times c$, where $h$, $w$, and $c$ denote the height, width, and channel counts, respectively. For both kinds of images, $h, w = 4096$, with $c = 1$ for HMI images and $c = 10$ for AIA images. The lower-left panel depicts predictors derived from SHARP parameters, which are summary statistics derived from HMI images. If the method performs classification, the outcome is the indicator of whether a flare will occur; if the method performs regression, the outcome is the future peak soft X-ray flux. The upper-right and lower-right panels show comparisons of actual values with mock predictions for classification and regression, respectively. In the upper right, the green curve represents mock flare probabilities; a classifier may output predicted classes or predicted class probabilities. The images and plots show data from around the time of an X8.7-class flare on 14 May 2024 in NOAA active region 13664 (HARP 11149).}
    \label{fig:schematic}
\end{figure}

\newpage

\subsection{Predictor-Outcome Combinations in Solar Flare Forecasting}
\label{subsec:types}

Different pairs of predictor and outcome types have been used in the solar flare forecasting literature. In Figure~\ref{fig:schematic}, we provide an overview of how predictions are produced by a machine learning method from different types of predictor data for both classification and regression in solar flare forecasting. Predictors are commonly computed from either the images of active regions or summary statistics on those regions. The upper left part of the schematic depicts predictors being computed from Helioseismic and Magnetic Imager (HMI) and (Atmospheric Imaging Assembly) AIA images, while the lower left part depicts predictors being computed from SHARP parameters, which are summary statistics calculated from HMI measurements. If the method performs classification, the outcome is a binary variable indicating whether a flare will occur; if the method performs regression, the outcome is a numeric variable, such as the future peak soft X-ray flux. The upper and lower right panels show comparisons of actual outcomes to mock predictions for classification and regression, respectively. The images and plots present data from around the time of an X8.7-class flare on 14 May 2024 in NOAA active region 13664 (HARP 11149). Here we briefly discuss some relevant works corresponding to the parts illustrated in Figure~\ref{fig:schematic}. Detailed numerical results will be presented in Sections~\ref{sec:Flare_Response} and~\ref{sect:pred_comparison}. Here, we briefly review the different predictor-outcome combinations in solar flare forecasting literature with machine learning methods.

\subsubsection{Image Predictors and Binary Responses} 

Image predictors typically consist of multi-channel AIA and HMI images of the Sun. Full-disk images of the Sun at their native resolution of $4096\times4096$ for each channel are usually too computationally and memory-intensive to serve as inputs to machine learning models, even when using batch learning. However, resizing these images through down-sampling or other interpolation methods can degrade spatial perception and erase meaningful, fine-grained details that are useful for flare prediction. On the other hand, multi-channel images focused on active regions~\citep{chen2024solar} must not only account for potential concurrent temporal effects (i.e., when active regions share timelines) but also for varying image resolutions, as active regions vary in size and shape. To address this, works such as \citet{Pandey2024} perform a rectangular crop that maximizes unsigned flux (a SHARP parameter). However, other work, such as \citet{sun2023tensor}, employs a similar yet distinct approach by performing a crop centered on the {polarity inversion line}~\citep[PIL,][]{Schrijver:2007}. PILs are typically the main source of flare activity, as shown in many works,  e.g., \citet{Zirin:1993, Green:2018, sun2021improved}, so such crops can not only keep predictive power high but also allow for succinct comparisons of features (such as topological and spatial features) that the model detects with the derived PIL for potential interpretability results. 

Multi-channel images of active regions are commonly adopted in flare prediction. Typically, machine learning models are trained to forecast whether a flare will be ``strong'' within a set time period: a binary prediction with response being $\geq \text{M}$ (or $\geq \text{C}$) flares. Despite its simplicity, binary classification still presents challenges such as class imbalance and ill-defined boundaries between flare categories, particularly near the chosen threshold (e.g., just below or above the M-class). Alternatively, models can try to predict the exact flare class, but this introduces similar boundary problems across multiple class thresholds. Many studies use image predictors to predict binary flare classes. We briefly review a subset of these. Focusing on works that use only 2D magnetogram imaging data, \citet{Park2018} trains a deep Convolutional Neural Network (CNN) on full-disk SDO/HMI magnetograms for binary predictions of C+ flares. Meanwhile, \citet{huang2018deep} and \citet{zheng2019solar} train CNNs to perform multi-class predictions of C-, M-, and X-class flares using {active region specific and} full-disk magnetograms {respectively}. On the other hand, \citet{li2022knowledge} and \citet{sun2022predicting} train CNNs on resized HMI magnetograms focused on active regions. Beyond using only 2D magnetogram data, \citet{jonas2018flare} applies Gabor filters and other techniques to a subset of AIA imaging channels to train a linear model to predict $\geq \text{M}$ binary flare outcomes. Furthermore, \citet{nishizuka2018deep} trains a {deep neural network (a MLP with skip connections called DeFN) on HMI and AIA imaging channels} to predict the binary responses for both $\geq \text{C}$ and $\geq \text{M}$ class flares. 
{Additionally \citet{nishizuka2021operational} later extends DeFN to also be trained with full-disk AIA images for binary predictions using both $\geq \text{C}$ and $\geq \text{M}$ decision rules. Another work that combines both the AIA and HMI images as predictors is \citet{sun2023tensor}, which proposes a tensor Gaussian process regression with contraction operators to combine AIA and HMI images to predict strong flares ($\geq$ M-class) versus weak flares ($\leq$ B-class).}

\subsubsection{Image Predictors and Continuous Responses}

Instead of predicting flare classes, another generalization of flare prediction is to directly predict the peak X-ray flux of a flare, treating the task as a regression problem rather than a classification problem. There is less work in this area because regression is inherently more difficult than classification, but we still provide a brief review here. 
In \citet{muranushi2015ufcorin}, the authors use 2D line-of-sight full-disk magnetogram data and then perform a wavelet transform to train a time-series regression model to predict the peak flux intensity of flares. On the other hand, \citet{sun2023tensor} uses both AIA and HMI imaging data aligned on active regions to perform a tensor regression to predict peak flux intensity. In these works, the models predict a continuous response, but the predictions are then converted into ``classification'' outputs using flare-class cutoffs to enable the use of commonly adopted classification metrics such as accuracy and skill scores. However, a work in which they do not bin their continuous predictions is \citet{vandersande2025thes}, which investigates key challenges and best practices to consider when treating flare prediction as a strictly regression problem.

\subsubsection{Vector Predictors and Binary Responses} 

In addition to high-resolution images that pose challenges for model training, many flare-forecasting studies rely on input data vectors derived from summary statistics or data-driven features computed from the images. Space-Weather HMI Active Region Patches (SHARPs, \citet{SHARPs}), derived from photospheric magnetic field data taken by the HMI aboard the SDO, provide 25 magnetic and geometric summary parameters. SHARPs are widely used for flare forecasting and have recently demonstrated success in SEP predictions~\citep{yu2025solar}. Similar to SHARPs, Space-Weather MDI Active Region Patches (SMARPs, \citet{SMARPs}) are derived from maps of the solar surface magnetic field taken by the Michelson Doppler Imager (MDI) aboard the Solar and Heliospheric Observatory (SOHO). \citet{sun2022predicting} adopts the SMARPs for flare forecasting. 

The binary classification of strong flares (such as $\geq$ M-class) against weak flares (such as $\leq C$ or $\leq B$ class) with multi-dimensional time series predictors is the most prevalent in the data-driven solar flare forecasting literature. 
Previous work on machine learning and deep learning models trained on multidimensional predictors extracted from photospheric magnetic field data (such as SHARPs) shows that carefully engineered vector physical summaries can already achieve competitive performance~\citep{bobra2015sola, florios2018forecasting, liu2019predicting, chen2019iden, nishizuka2018deep,Wang2020lstm,nishizuka2021operational}. Vector predictors are not limited to features derived from images. \citet{verma2026optimal} studies the probability of a strong flare event as an extreme-event prediction using soft X-ray flux time-series vectors.

\subsubsection{Vector Predictors and Continuous Responses}

Flares are classified based on their peak intensity, which is defined to be the peak value of the GOES X-ray flux during the flare. For example, M-class flares have a peak intensity that is at least $\SI{e-5}{W.m^{-2}}$ and less than $\SI{e-4}{W.m^{-2}}$, and X-class flares have a peak intensity that is at least $\SI{e-4}{W.m^{-2}}$ \citep[][]{goes8-15_readme}. A typical formulation of the flare forecasting problem is predicting whether an M+ flare will occur in a future 24-hour period \citep[][]{leka2018sola}. In this formulation, the outcome is binary. Table~\ref{tab:max_flux_vs_flare} provides a justification for an alternative formulation in which the outcome is continuous. The table is based on the Space Weather Analytics for Solar Flares (SWAN-SF) benchmark dataset, which contains various kinds of data, including flare and GOES X-ray flux data, for the period 2010-2018 \citep[][]{angryk2020multivariate}. For each date between May 1, 2010 and December 29, 2018, we checked whether an M+ flare occurred and whether the maximum X-ray flux was at or above the M-class threshold of $\SI{e-5}{W.m^{-2}}$. All but four dates either had both an M+ flare and an M+ maximum X-ray flux or had neither. This shows that predicting whether an M+ flare will occur during a 24-hour window is essentially equivalent to predicting whether the maximum X-ray flux during the window will be above the M-class threshold. Note that this observation is not trivial since a 24-hour window can contain the rising phase of an M+ flare without having an exceedance of the M-class threshold or can follow the decaying phase of an M+ flare yet still have an exceedance of the threshold. Threshold exceedance alone does not determine whether an M+ flare is occurring; several other criteria must be satisfied, as outlined in Appendix A of \citet{goesr_l2_userguide}. A threshold exceedance can be predicted by forecasting the maximum X-ray flux, which is a continuous outcome. See \citet{verma2026optimal} for more discussions. 

\begin{table}[ht]
    \centering
    \begin{tabular}{lcc}
    \toprule
    & \multicolumn{2}{c}{\textbf{Date Had M+ Flare?}} \\
    \cmidrule(r){2-3}
    \textbf{Date Had M+ Maximum X-Ray Flux?} & \textbf{Yes} & \textbf{No}        \\
    \midrule
    \textbf{Yes} & 434 (17\%) & 0 (0\%) \\
    \textbf{No} & 4 (0.15\%) & 2,146 (83\%) \\
    \bottomrule
    \end{tabular}
    \caption{The joint distribution of the M+ maximum flux and flare indicators for dates between May 1, 2010 and December 29, 2018, inclusive. The SWAN-SF benchmark dataset described in \citet{angryk2020multivariate} was used to compute the counts.}
        \label{tab:max_flux_vs_flare}
\end{table}

Forecasting a continuous outcome has several advantages over forecasting a binary outcome. First, a prediction of a binary outcome is a prediction of whether an event of interest will occur, but not of its severity if it does. A prediction of a continuous outcome involves predictions of both occurrence and severity~\citep[see e.g.,][]{jiao2020solar,verma2026optimal}. Second, the flare class thresholds are arbitrary, meaning that there is no substantive difference between the strongest flares in one class and the weakest flares in the next higher class. If ignored, the lack of a clear distinction between adjacent flare classes can make it difficult for a binary forecasting method to perform well because the method may have trouble correctly predicting flares near a class boundary. Several studies use a continuous outcome rather than a binary outcome. For example, \citet{vandersande2025thes} model the maximum soft X-ray flux in a 24-hour window using linear regression, random forests, multilayer perceptrons (MLPs), and convolutional neural networks (CNNs). \citet{boucheron2015pred} employs support vector regression (SVR) trained on magnetic features extracted from MDI images, using a linearly spaced continuous-valued label vector to represent flare size. \citet{jiao2020solar} develops a mixed LSTM model to forecast the maximum flare intensity, and considers classification models built on top of the regression.

\subsection{Outline}

In the sequel, we will describe the issues that arise when using the various types of data mentioned above in Section~\ref{subsec:types} and suggest ways to address them. In Section~\ref{sec:Flare_Response}, we focus on response data, covering various data sources, quality, and a systematic comparison. In Section~\ref{sect:pred_quality}, we focus on predictor data, with image-based predictors discussed in Section~\ref{subsect:image_preds} and vector predictors discussed in Section~\ref{subsect:SHARPs_SMARPs}.  Finally, in Section~\ref{sect:pred_comparison}, we show how predictive performance can be degraded if the issues we raise aren't addressed. We conclude in Section~\ref{sec:conclusion} with take-home messages on our numerical comparison results, clarifying strengths and limitations of different data products.


\section{Flare Response: Data Source, Quality, and Processing}
\label{sec:Flare_Response}

\subsection{Sources of Data Products for Flare Events}\label{subsect:Organizations}

\subsubsection{GOES X-ray Flare Data}
Geostationary Operational Environmental Satellites (GOES) are operated by NOAA (National Oceanic and Atmospheric Administration
), and observe the Sun in X-ray spectrum. GOES flare event data, 1-minute average X-Ray flux data, are  one of the most commonly used resources in flare forecasting tasks and is available from multiple organizations. The GOES X-Ray Sensor (XRS) measurements have been a crucial component of space weather operations since 1975. XRS measurements are in two bandpass channels commonly referred to as the XRS-A (0.05-0.4 nm, {known as the hard X-ray band}) and XRS-B (0.1-0.8 nm, {known as the soft X-ray band}). 

A sudden yet persistent increase in flux is detected as a likely flare. For GOES-R (GOES 16-18) data, the detection algorithm uses a variety of criteria, including an exponential fit, to determine that a flare has begun. Specifically, the algorithm collects a small block of 1-minute averaged XRS-B flux data and smooths the data by a small boxcar filter. The algorithm then checks whether the flux shows signs of a significant rise by evaluating two key features: the presence of an inflection point, where the curve transitions to a faster upward slope, and whether the rise exceeds a defined threshold based on the standard deviation of earlier values. If these conditions are met, it fits the smoothed data to an exponential function. The exponential curve must also show a strong, concave upward shape based on the configurable parameters. If all these criteria are satisfied, the algorithm declares a flare start. The flare peak time is recorded when the flux reaches its maximum. The end time of a flare is declared when the X-ray flux has declined to halfway between the flare peak and the background level~\citep{goesr_l2_userguide}. 

\begin{figure}[!ht]
    \centering
    \includegraphics[width=\textwidth]{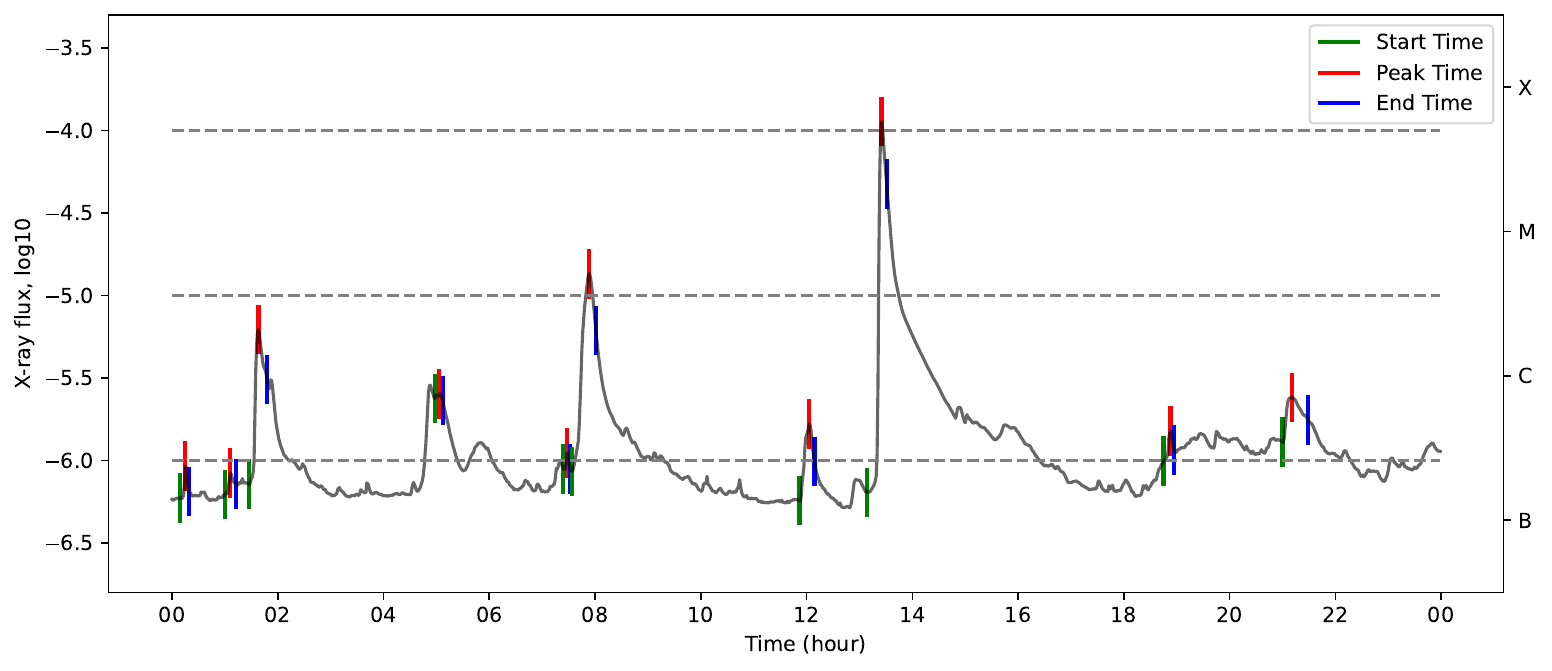}
    \caption{The 1-minute averaged X-Ray flux in log10 scale and the flare events at May 3rd, 2022. The short, vertical lines represent the event time of flares. Not every local maximum in flux will be labeled as one flare event. The flux and the flare events are the Science-Quality data processed by the NOAA National Centers for Environmental Information (NCEI).}
    \label{fig:flux_and_flares_20220503}
\end{figure}

The magnitude of a flare is defined by NOAA Space Weather Prediction Center (SWPC) with a flare index that is based on the 1-minute XRS-B irradiance at the peak of the flare. Flare indices are denoted by a letter and a number based on the log 10 peak irradiance of the flare in watts per square metre (W/$\text{m}^2$) (X: $10^{-4}$ W/$\text{m}^2$, M: $10^{-5}$ W/$\text{m}^2$, C: $10^{-6}$ W/$\text{m}^2$, B: $10^{-7}$ W/$\text{m}^2$, and A: $10^{-8}$ W/$\text{m}^2$). Figure~\ref{fig:flux_and_flares_20220503} shows the 1-minute averaged X-Ray flux at a log10 scale and the flare events on May 3rd, 2022. There are 1 X-class flare, 1 M-class flare, 5 C-class flares, and 3 B-class flares. 

A notable difference between the GOES-R (GOES 16-18) and previous GOES data is that the operational (near-real-time) data prior to GOES-16 had scaling factors applied by the SWPC (Space Weather Prediction Center) to adjust the GOES 8-15 irradiance to match fluxes from GOES-7. Thus, the flare index before GOES-16 was based on the operational irradiance with scaling factors. To obtain the true irradiance, the scaling factor of 0.7 (for the XRS-B channel) have to be removed (see Section~\ref{subsect:sci_vs_oper}).

Operational flare data based on the 1-minute X-Ray flux is prepared and stored by the SWPC. SWPC provides near-real-time and recently observed X-ray flux data, solar and geomagnetic indices, and solar event reports created from preliminary reports. Specifically, SWPC offers Solar Events Lists (SEL) from 1996 to the present, accessible through its public warehouse\footnote{see \url{ftp://ftp.swpc.noaa.gov/pub/warehouse/}}. The SEL provides daily preliminary lists of solar events, including bright surges on the limb, filament disappearances, H-alpha optical flares, X-ray flares, and others. By selecting the \textit{XRA} event type, users can access operational X-ray flare event data dating back to 1996~\citep{SEL}. We refer to the flare event lists from the SEL as the \texttt{SWPC-FTP} list. The \texttt{SWPC-FTP} list includes flare times, magnitudes, and the assigned NOAA solar active region numbers. Additionally, the warehouse contains the daily SRS (Solar Region Summary) report, which provides detailed information about the solar active regions, including their positions.

Science-Quality XRS data is produced by NOAA's National Center for Environmental Information (NCEI) and differs from the operational products used at SWPC in that the data have been preprocessed and incorporate retrospective fixes for issues and outages in the operational product. Specifically, NCEI has reprocessed GOES 8-18 operational flare summary data to produce science-quality flare summary data (see Section~\ref{subsect:sci_vs_oper}). The GOES 1-7 data will be completed in 2026. The Science-Quality flare summary data are based on 1-minute Science-Quality X-ray irradiance data. We refer to this flare catalog as the \texttt{Science-Quality} list in this paper. The \texttt{Science-Quality} list does not include the corresponding active region (AR) number for each flare event. However, NCEI provides Science-Quality flare location information for GOES-R (GOES 16-18) data. The flare location on the solar disk is determined based on the measurements from the high flux XRS-B2 quad diode detector and was calibrated and validated with comparisons to flare locations from the {Heliophysics Event Registry}\footnote{\url{https://www.lmsal.com/heksearch/}, \citet{goesr_l2_userguide}}. And this can be used to match flares to the corresponding ARs. Table~\ref{tab:xray-flare-sources} summarizes the different GOES X-ray flare sources.

\begin{table}[!ht]
\centering
\caption{Overview of GOES X-ray Flare Sources. }
\label{tab:xray-flare-sources}
\begin{tabular}{lccccc}
\toprule
\textbf{Reference} & \textbf{Time Range} & \textbf{Satellite} & \textbf{Category} & \textbf{AR} & \textbf{Location} \\
\midrule
\href{ftp://ftp.swpc.noaa.gov/pub/warehouse/}{\texttt{SWPC-FTP}}        & 1996/07/31 -- present    & GOES 7--16  & Operational& Yes & No \\
\href{https://www.ncei.noaa.gov/products/satellite/goes-r}{\texttt{Sci-Quality}}     & 2009/09/01 -- 2020/03/04 & GOES 8--15  & Sci-Quality& No  & No \\
\href{https://www.ncei.noaa.gov/products/goes-r-extreme-ultraviolet-xray-irradiance}{\texttt{Sci-Quality}}     & 2017/02/09 -- present    & GOES 16--18 & Sci-Quality& No  & Yes \\
\href{https://www.lmsal.com/heksearch/}{\texttt{SunPy-HEK}}       & 1975 -- present          & GOES 1--16  & /     & Yes & Yes \\
\bottomrule
\end{tabular}
\end{table}

\subsubsection{AIA Flare Catalog, SolarSoft Latest Events, and HEK Repository}
Besides GOES X-Ray flare, AIA flares are solar flares detected and characterized based on extreme ultraviolet (EUV) observations from the AIA~\citep{lemen2012atmospheric} aboard SDO~\citep{pesnell2012solar}. 
The detection process uses a peak-detection algorithm applied to binned AIA images, primarily at 193\AA{}, to identify the start, peak, and end times and the approximate location of the flare. The AIA flare detection algorithm detects flares as intervals of positive derivative and negative derivative around a local maximum~\citep{Martens2012}.

SolarSoft Latest Events {(hereafter referred to as the \texttt{SSW} flare catalog; \citealt{Freeland2002})} is another catalog that regularly detects and collects solar flare events. The catalog is primarily designed to assign spatial locations to GOES flare events using the best available full-disk X-ray/EUV data. Since 2010, AIA 94\AA{} has been the primary data source, and currently, GOES-R series SUVI 93/94\AA{} serves as a backup when AIA data is unavailable. It is independent of specific instruments.

The HEK (Heliophysics Events Knowledgebase) is a comprehensive metadata system designed to catalog and provide access to solar events, facilitating efficient data retrieval for researchers~\citep{hek}. The HEK serves as an integrated event repository rather than a direct data source. It aggregates flare information from multiple sources. The HEK incorporates AIA flare detections from automated feature-finding tools applied to 94 \AA{} EUV images. It also ingests external sources, including the SWPC operational GOES X-Ray flare list and the SSW catalog.


The {\texttt{SunPy}~\citep{sunpy20}} package provides access to solar event data through its interface to the HEK.  According to the official \texttt{SunPy} documentation and the open-source code, users can retrieve the desired SWPC operational flare events from the HEK repository by specifying the appropriate event type and observatory. In this work, we refer to the resulting dataset as the \texttt{SunPy-HEK} GOES flare list. However, this list exhibits several discrepancies compared to the flare list available in the SWPC open warehouse, particularly in event counts and Active Region (AR) number information. Between January 1, 1997, and July 21, 2024, 289 flares (1.14\%) appear in the \texttt{SunPy-HEK} list but are absent from the \texttt{SWPC-FTP} list, while 817 flares (3.15\%) recorded by SWPC are not found in the \texttt{SunPy-HEK} list. Moreover, 9842 flare events have differing AR numbers between the two lists. 

\begin{figure}[!htb]
    \centering
    \includegraphics[width=\textwidth]{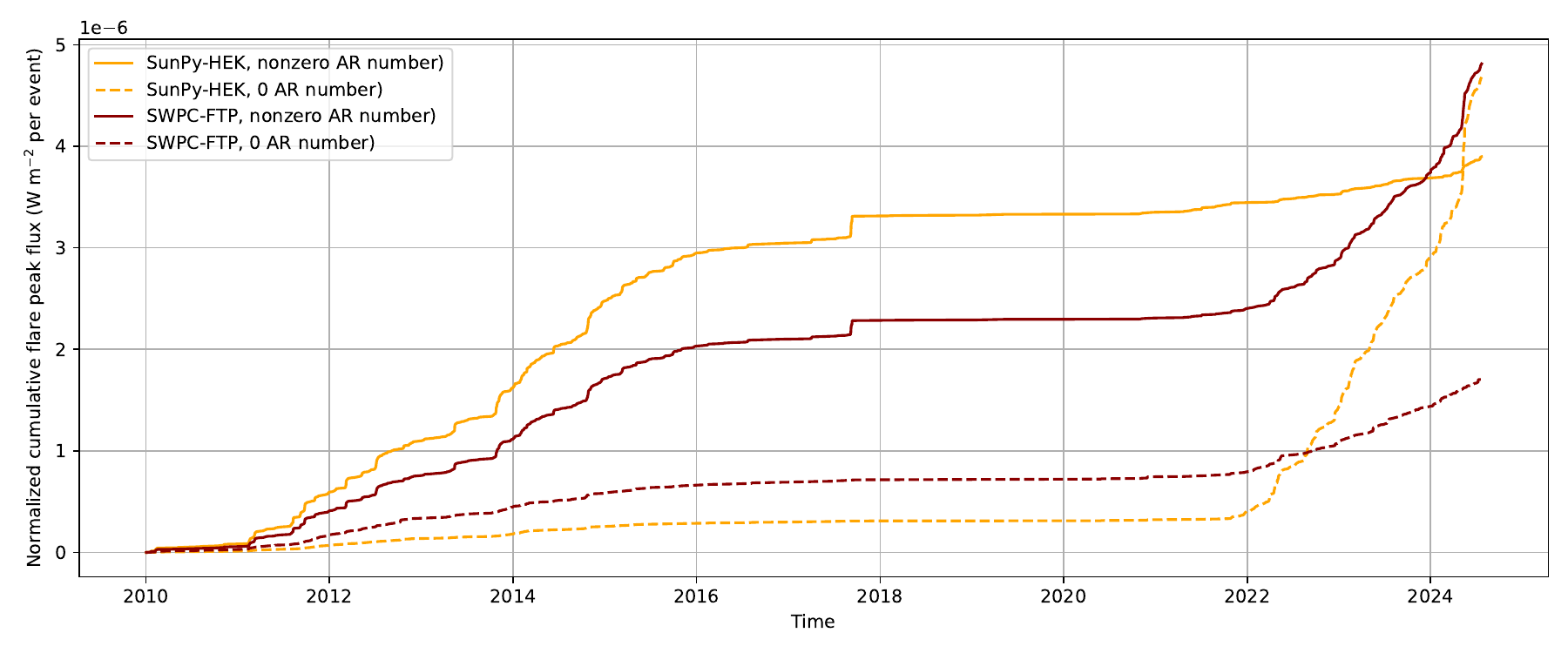}
    \caption{{Normalized cumulative flare peak flux as a function of time for \texttt{SunPy-HEK} and \texttt{SWPC-FTP} data sources, separated by whether a nonzero NOAA AR number is assigned. For each category, the cumulative peak flux is normalized by the total number of events to account for differences in sample size. Before 2020, the curves from different sources and AR assignments are broadly consistent. After 2020, the \texttt{SunPy-HEK} events with AR number equal to zero exhibit a suspicious increase in normalized cumulative intensity because more than 3000 flares are mislabeled with AR number 0.}}
    \label{fig:sunpyFTP_cum_intensity}
\end{figure}

Figure~\ref{fig:sunpyFTP_cum_intensity} shows the normalized cumulative flare intensity since January 1, 2010, for both the \texttt{SunPy-HEK} and \texttt{SWPC-FTP} flare lists. A notable disparity appears after 2022, due to more than 3,000 flares being mislabeled with AR number 0 in the \texttt{SunPy-HEK} list. The HEK operation team suggested that the difference in event counts may be due to the double retrieval of the same flare after SWPC revisions. Thus, we do not recommend using the \texttt{SunPy-HEK} GOES flare list, especially the GOES-R data, for research purposes.

\subsection{Comparison of Science Quality and Operational Flare Response}\label{subsect:sci_vs_oper}

NCEI produces and restores science-quality \href{https://www.ncei.noaa.gov/products/goes-r-extreme-ultraviolet-xray-irradiance}{GOES XRS L1b and L2 datasets} in \texttt{netCDF} format. The science-quality data have been reprocessed from the start of the mission to the present date and incorporate retrospective fixes for issues and outages in the operational product. The XRS L1b data contain high-cadence soft X-ray irradiance measurements, and the L2 data consist of higher-order products such as irradiance averages, flare event summaries, and flare location products. Science-quality L2 data products are created from science-quality L1b data. Up to October 2025, science-quality flare summary products are available for GOES-R (GOES 16, 17, 18) and GOES 8-15 satellites, dating from 1995/01/03 to the present. The 1-minute flux average product, another commonly used L2 product, has been processed to science-quality for GOES 8-18 satellites. This paper focuses primarily on the flare summary product, which contains the most widely used flare-response information for flare-prediction tasks.

The science-quality XRS data have been reprocessed with numerous corrections, including smoothing, accuracy, scaling factor consistency, bandpass corrections, and improvements to the data quality flag~\citep{goes8-15_readme}. The SWPC scaling factor has the most crucial effect on the flare summary product. Science-quality XRS irradiances are provided in physical units of W/m$^2$, while the operational data before GOES-16 had SWPC-adjusted scaling factors that adjusted GOES 8–15 irradiance to match those of GOES 7. Since the flare summary product is based on the L1b data using the flare detection algorithm, the threshold for each flare class increased by 42\%, meaning that an individual solar flare before GOES-R needed to be 42\% larger than a flare after GOES-R to be given the same level of magnitude. Thus, fewer flares were covered in the operational data set during GOES 8-15. The science-quality data were corrected to exclude the SWPC scaling factors.

\begin{figure}[!htb]
    \centering
    \includegraphics[width=0.95\textwidth]{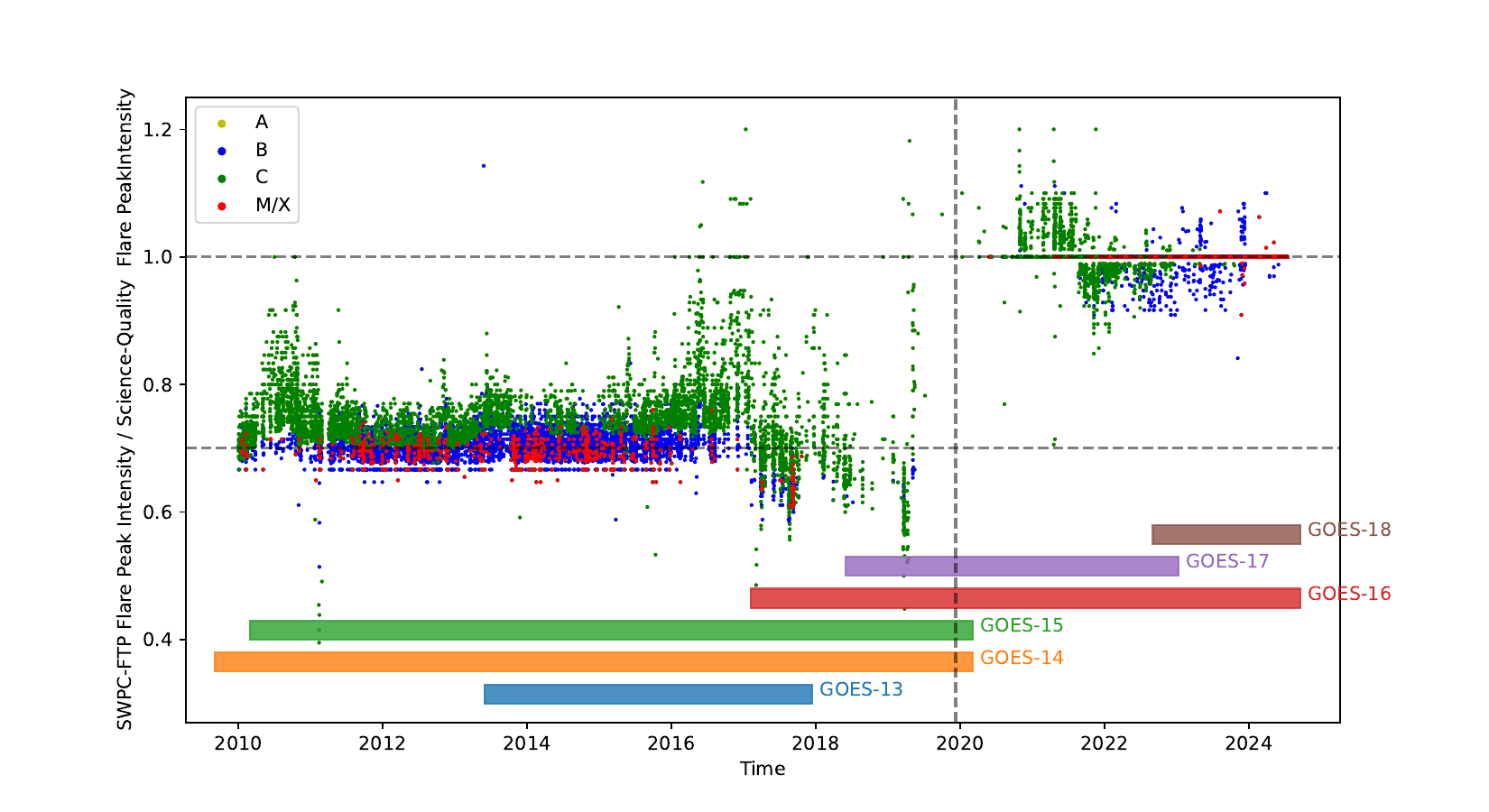}
    \caption{The flare peak flux ratio of \texttt{SWPC-FTP} list to the NCEI science-quality list through 2010/01/01 - 2024/7/21. {Each point shows the ratio of \texttt{SWPC-FTP} (operational) to \texttt{Sience-Quality} flare peak intensity for matched events. For each operational flare, a corresponding NCEI science-quality flare is identified by matching peak times within $\pm 2$ minutes; the ratio is then computed as the operational flare intensity divided by the science-quality flare intensity. Events without a match are excluded.}}
    \label{fig:ratio}
\end{figure}

Figure~\ref{fig:ratio} shows the flare magnitude inconsistency in the operational data. {Before December 2019, the SWPC applied a rescaling factor so that the ratio of \texttt{SWPC-FTP} flare intensity and \texttt{Science-Quality} flare intensity was centered at 0.7, whereas the centered ratio returned to 1 as GOES-16 became the primary operational satellite. The colored rectangles in Figure~\ref{fig:ratio} show the data availability time ranges for GOES 13-18, which differ from the serving time, as the primary satellite providing operational data. GOES-16 became the primary geostationary satellite in December 2019.}

\begin{figure}[!htb]
    \centering
    \includegraphics[width=\textwidth]{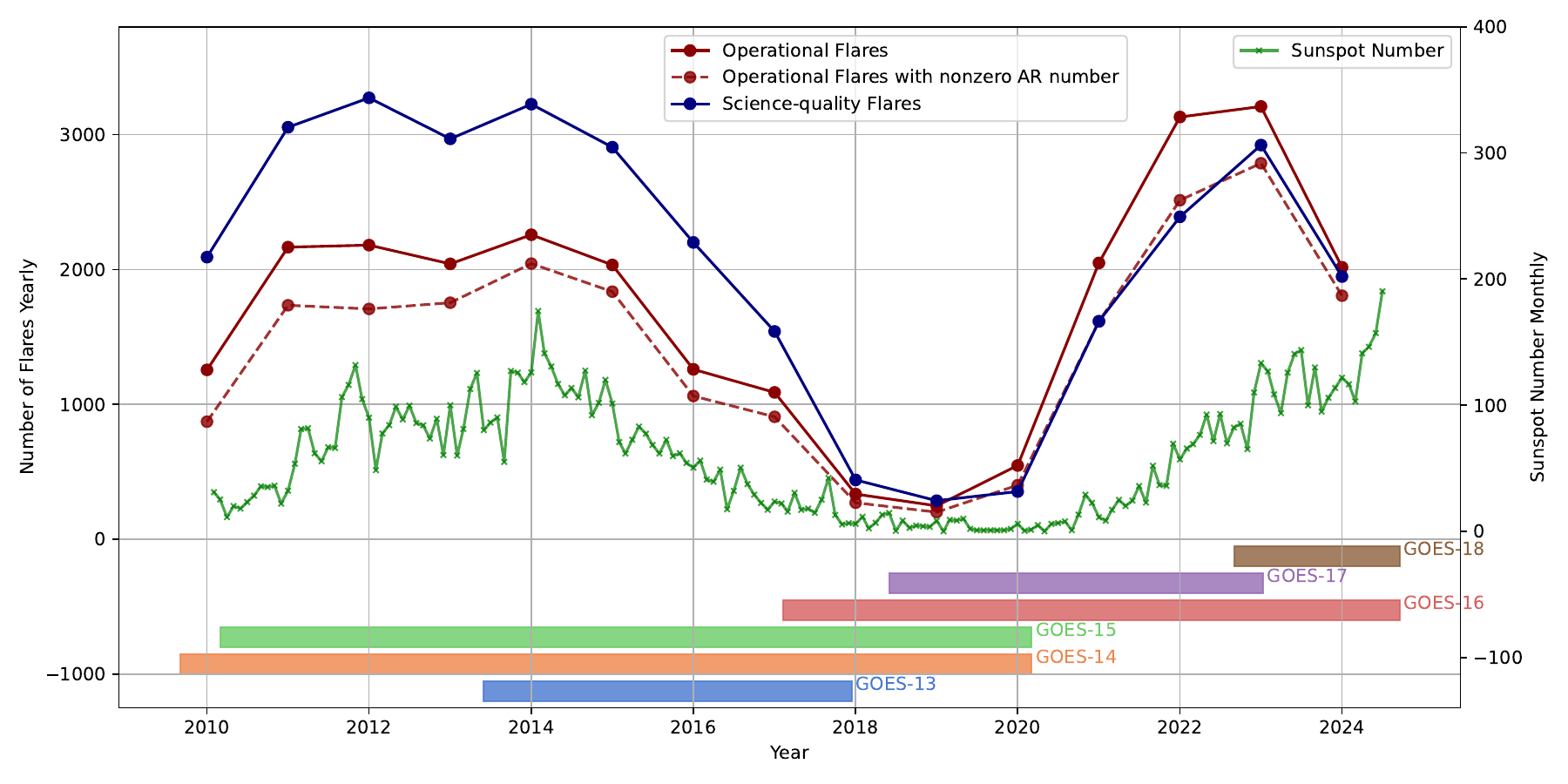}
    \caption{Flare event numbers from NCEI \texttt{Science-Quality} flare data and \texttt{SWPC-FTP} flare data. The dashed line represents \texttt{SWPC-FTP} flare records with a nonzero AR number. }
    \label{fig:counts}
\end{figure}

Moreover, the SWPC rescaling factor contributes to discrepancies in flare counts across different catalogs. Figure~\ref{fig:counts} compares the number of events in the operational and science-quality datasets. {Prior to December 2019, fewer flares were recorded in the operational data due to the SWPC scaling factor. After the transition to primarily using GOES-16, the number of science-quality flare events is smaller due to the correction.}

\subsection{Cross-Comparison of Science-Quality, Operational, and SSW Flare Lists}\label{subsect:cross_checking}

To further investigate the differences of these flare lists, we compared the \texttt{Science-Quality}, \texttt{SWPC-FTP}, and \texttt{SSW} flare lists. The \texttt{Science-Quality} catalog was obtained through the NCEI website. The \texttt{SWPC-FTP} catalog is directly downloaded from the SWPC open warehouse due to the inconsistency between the FTP archive and \texttt{SunPy-HEK} records. The \texttt{SSW} flares were downloaded from their web archive~\citep{Freeland2002}. In addition, we downloaded the Science-Quality flare location dataset from NCEI and augmented the \texttt{Science-Quality} list with pinpoint location information. Finally, \texttt{Science-Quality} and \texttt{SSW} have pinpoint flare location coordinates, while \texttt{SWPC-FTP} and \texttt{SSW} have NOAA AR numbers. 

The period of interest is from 2010-01-01 to 2024-12-31. NCEI provides processed science-quality data for each instrument separately, and the data availability coverage of one instrument usually does not align with the time range during which the instrument is in operational use. For example, although GOES-16 reached geostationary orbit on 2016-11-29 and began providing flare event data on 2017-02-09, it did not become NOAA’s primary source of solar X-ray flux until 2019-12-09. GOES-16 was subsequently replaced by GOES-18 as the primary satellite in December 2024.

Consequently, over the study period (2010-01-01 to 2024-12-31), multiple satellites (GOES-13, -14, -15, and -16) served as the operational primary data source at different times. To ensure consistency with the operational flare catalog, we construct the \texttt{Science-Quality} flare list by combining NCEI science-quality flare summary products from different instruments, using time intervals aligned with the periods during which each instrument serves as the operational primary satellite. Because flare location products are available only for the GOES-R series and such geographic information is essential for training predictive models, we include the full GOES-16 time range starting from 2017-02-19. Prior to GOES-16, science-quality data are compiled from earlier GOES instruments based on their periods of primary operational use~\citep{goes1-15_readme}.

Table~\ref{tab:goes_intruments} summarizes both the timeline used to construct the science-quality flare dataset and the transmission periods of GOES instruments. {The ``\texttt{Science-Quality}'' column represents the time intervals over which data from each GOES instrument are selected and combined in this study. These intervals are chosen to align with the periods during which each instrument serves as the operational primary satellite, as indicated in the “Primary” column. The ``Primary'' and ``Secondary'' columns denote the satellites as the operational sources of solar X-ray flux at a given time. In general, there have been two operational GOES satellites for each instrument, a primary and a secondary. Primary satellite data should be used in preference to secondary satellite data whenever available. There will be gaps even in good data for the primary satellites due to eclipses and in-flight calibrations. }

\begin{table}[!htb]
    \centering
\caption{{Timeline of the compiled science-quality flare data list and GOES satellite transitions. The ``\texttt{Science-Quality}'' column denotes the time intervals over which data from each GOES instrument are used to construct the combined science-quality flare dataset in this study (see Section~\ref{subsect:cross_checking} for details). Satellites are listed in the table only when changes occur. Numbers (e.g., 13–18) refer to GOES satellite identifiers.} }
    \label{tab:goes_intruments}
    \begin{tabular}{|c|c|c|c|} \hline 
         Date&  Science-Quality& Primary & Secondary\\ \hline 
 2023-01-04& & & 18\\ \hline 
 2021-08-24& & & 17\\ \hline 
 2019-12-09& & 16& 15\\ \hline 
 2017-12-12& & & 14\\ \hline 
 2017-08-23& & 15& 13\\ \hline 
 2017-08-16& & & 14\\ \hline 
 2017-08-08& & 13& 15\\ \hline 
 2017-02-07& 16& & \\ \hline 
 2016-06-09& 15 & 15& 13\\ \hline 
 2016-05-16& & & 15\\ \hline 
 2016-05-12& 14 & 14& 13\\ \hline 
 2015-06-09& 13 & 13& 14\\ \hline 
 2015-06-09& 15& 15& \\ \hline 
 2015-05-21& 14 & 14& \\\hline 
 2015-01-26&  &  & 13\\ \hline 
 2012-11-19&  15&  15& 14\\ \hline 
 2012-10-23&  14&  14& 15\\ \hline 
 2011-09-01&  &  & 14\\ \hline 
 2010-10-28& 15& 15& None\\ \hline 
 2010-09-01&  &  & 15 \\ \hline 
 2009-12-31&  14&  14& None\\ \hline
    \end{tabular}
\end{table}

Finally, after compiling data across instruments over this time range, we obtain 33,215 \texttt{Science-Quality} records, including 139 X-class flares and 2,497 M-class flares. We also downloaded 27,586 \texttt{SWPC-FTP} and 33,489 \texttt{SSW} flares. We perform minimal data cleaning on these lists, only ensuring the validity of start time, peak time, and flare magnitude features. Table~\ref{tab: flare-counts} presents the number of flares with different magnitudes in our compiled  \texttt{Science-Quality}, \texttt{SWPC-FTP}, and \texttt{SSW} lists.

\begin{table}[!ht]
\centering
\caption{Summary of flare lists from \texttt{Science-Quality}, \texttt{SWPC-FTP} and \texttt{SSW}. The data range is from 2010-01-01 to 2024-12-31.}
\label{tab: flare-counts}
\begin{tabular}{lcccccc}
\toprule
\textbf{Source} & \textbf{X} & \textbf{M} & \textbf{C} & \textbf{B} & \textbf{A} & \textbf{Total} \\ 
\midrule
\texttt{Science-Quality} & 139 & 2497 & 19207 & 11372 & 0 & 33,215 \\
\texttt{SWPC-FTP} & 126 & 2193 & 16049 & 9120 & 98 & 27,586 \\
\texttt{SSW} & 124 & 2194 & 19791 & 10023 & 1357 & 33,489 \\
\bottomrule
\end{tabular}
\end{table}

To cross-compare different flare catalogs, we then attempt to match each \texttt{Science-Quality} flare with corresponding events in the \texttt{SWPC-FTP} and \texttt{SSW} catalogs based on temporal characteristics (i.e., flare peak time) and flare peak intensity. Specifically, we define matching tolerances for both quantities. For a given flare with peak time $t$ and intensity $I$, we first identify candidate events in the comparison catalog whose peak times fall within a predefined time tolerance interval, i.e., $|t-t'|\leq \Delta_t$, where $\Delta_t$ denotes the time tolerance.  Among those candidates, we further require that the absolute difference in log\textsubscript{10} intensity satisfies $|\log_{10}(I) - \log_{10}(I')| \leq \Delta_I$,  where $\Delta_I$ is the intensity tolerance. If multiple candidates satisfy both criteria, the match is determined by selecting the flare with the smallest peak time difference. If no candidate satisfies both criteria, the flare is treated as unmatched.

\begin{figure}[!htb]
    \centering
    \includegraphics[width=\textwidth]{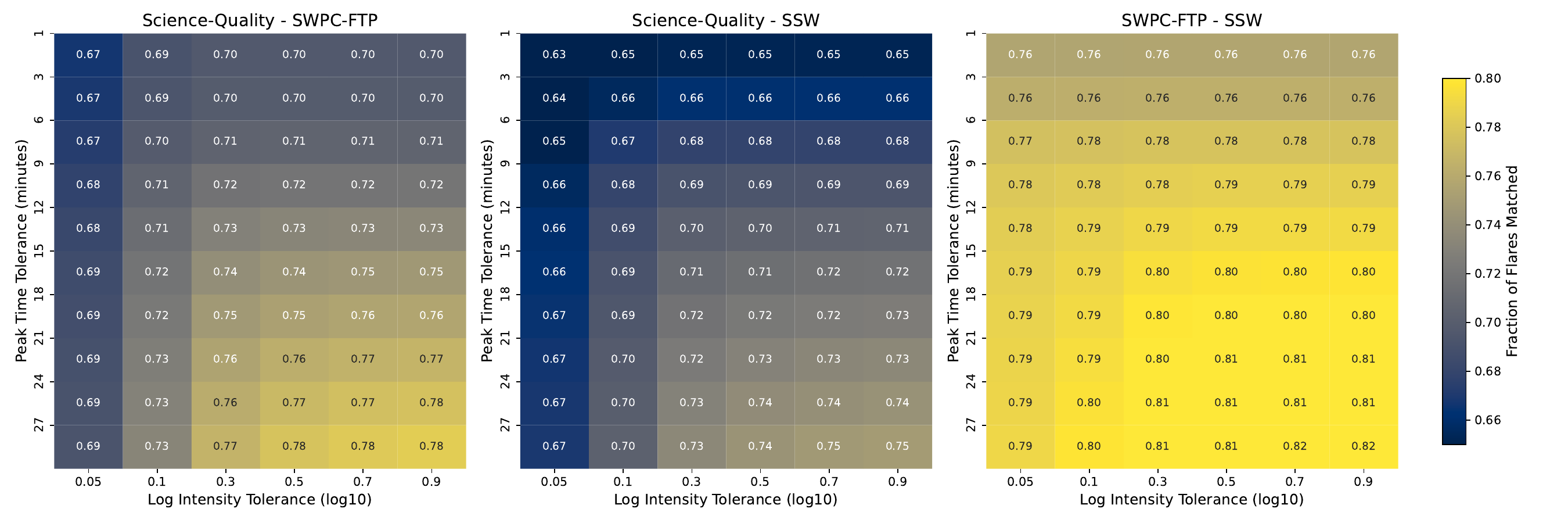}
    \caption{Matched flare fraction under different peak-time and intensity tolerances. {Matching is performed over a grid of peak time tolerances (1-27 minutes) and log\textsubscript{10} intensity tolerances (0.05--0.9). For each tolerance pair, candidate matches are first selected based on peak time proximity and then filtered by intensity similarity. Each cell shows the fraction of flares in the first catalog listed in the panel title that are successfully matched to the second catalog.}  }
    \label{fig:matching_heatmap}
\end{figure}

{To evaluate the impact of matching tolerances, we perform the matching procedure over a grid of peak-time and intensity tolerances. Figure~\ref{fig:matching_heatmap} shows the fraction of matched flares under various tolerance settings. Specifically, the peak time tolerance $\Delta_t$ ranges from 1 to 27 minutes, and the log\textsubscript{10} intensity $\Delta_I$ ranges from 0.05 to 0.9. For each pair $(\Delta_t, \Delta_I)$, we match flares using the criteria described above and compute the fraction of flares that are successfully matched in the comparison catalog. }

One important consideration is that, before the designation of GOES-16 as the primary satellite, flare intensities reported in the \texttt{SWPC-FTP} catalog were systematically affected by the SWPC scaling factor. This results in reported $\log_{10}$ intensities that are approximately 0.15 ($-\log_{10}(0.7)$) lower than their true values. The extent to which the \texttt{SSW} catalog inherits this bias is not explicitly documented. However, our analysis shows that the \texttt{SSW} catalog exhibits greater similarity to \texttt{SWPC-FTP} than to any other pair of catalogs. Specifically, we first matched flares between the \texttt{Science-Quality} and \texttt{SSW} lists using only peak time with a strict tolerance of zero. For these temporally matched events, we computed the $\log_{10}$ intensity differences (in absolute value) across two periods: before 2019-12-07 (prior to GOES-16 availability) and after 2019-12-09 (once GOES-16 data were adopted operationally). 

\begin{figure}[htb]
    \centering
    \includegraphics[width=0.8\textwidth]{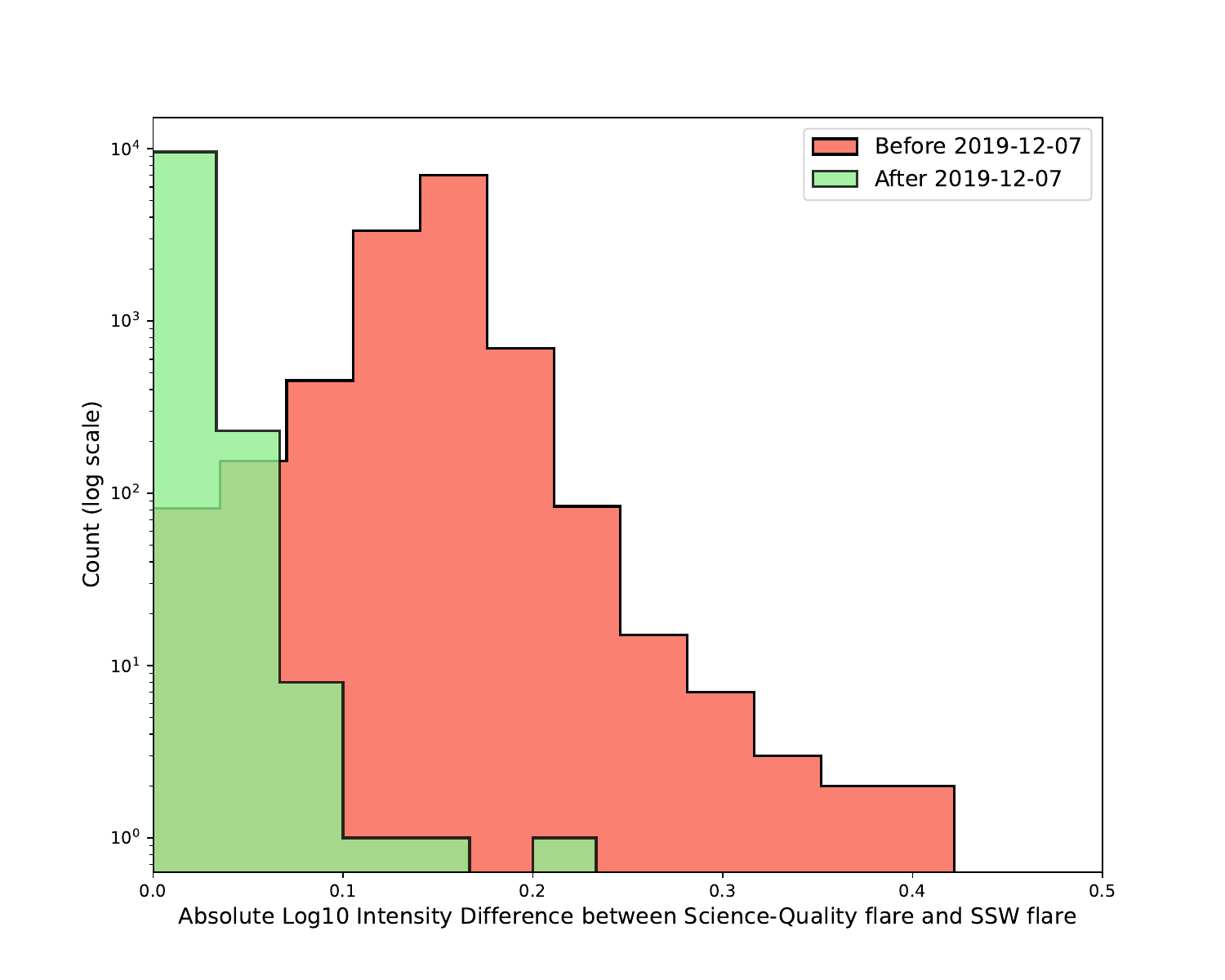}
    \caption{Distribution of absolute $\log_{10}$-intensity differences between matched \texttt{SWPC-FTP} and \texttt{SSW} flares. For each flare in the \texttt{Science-Quality} catalog, the corresponding \texttt{SSW} flare is identified by requiring the same peak time. Before the operational start of GOES-16 (2019-12-07), the differences exhibit a clear mode around 0.15, approximately $-\log_{10}(0.7)$, indicating that the \texttt{SSW} catalog retains the SWPC scaling factor during this period.}
    \label{fig:sci_ssw_peak0_loose_intensity_compare}
\end{figure}

Figure~\ref{fig:sci_ssw_peak0_loose_intensity_compare} shows the resulting histograms. Before GOES-16 became operational, the absolute intensity differences clustered around 0.15, consistent with the expected offset introduced by the SWPC scaling factor ($-\log_{10}(0.7)$). After 2019-12-07, the differences are centered near zero, indicating that this bias has been removed. Together with the matching-proportion pattern shown in Figure~\ref{fig:matching_heatmap}, these results demonstrate that the \texttt{SSW} catalog is not free from the SWPC scaling prior to GOES-16 and, more generally, that \texttt{SSW} aligns more closely with \texttt{SWPC-FTP} than with the \texttt{Science-Quality} catalog. Therefore, the \texttt{SSW} list should be treated as an operational dataset. Consequently, for events occurring before 2019-12-09, we adjusted the flare log intensity for \texttt{SWPC-FTP} and \texttt{SSW} by removing the SWPC scaling factor in log scale.

\begin{figure}[htb]
    \centering
    \includegraphics[width=0.75\textwidth]{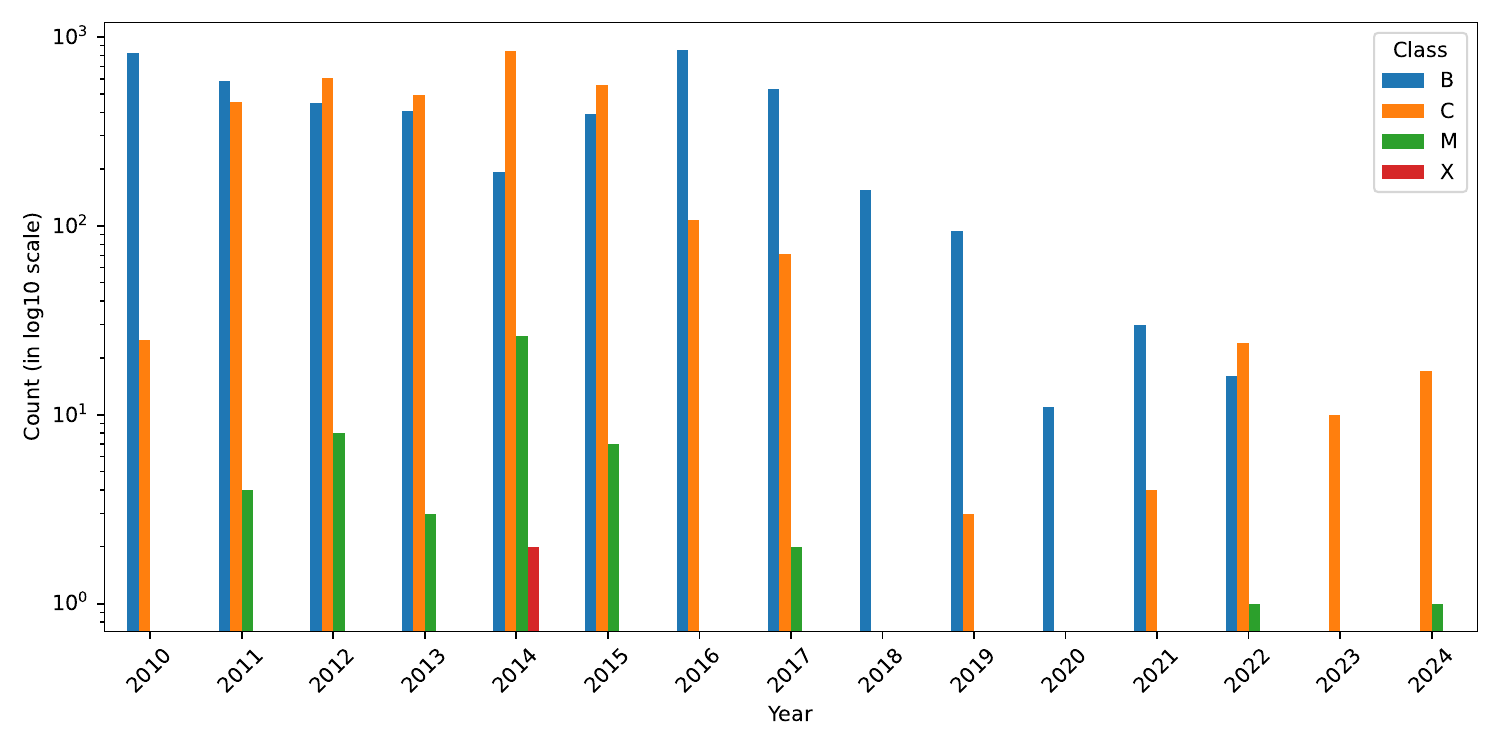}
    \caption{Number of solar flare events of each class in years 2010-2024 from \texttt{Science-Quality} flare list that are not matched to any flare event in the SWPC-FTP or SSW catalogs.}
    \label{fig:not_matched_sci}
\end{figure}

In this study, we adopt a 15-minute peak-time tolerance and a 0.3 $\log_{10}$ intensity tolerance as criteria for identifying corresponding flare events across different catalogs. As shown in Figure~\ref{fig:matching_heatmap}, when the log intensity tolerance is set over 0.3, the fraction of matched flares remains relatively stable. Figure~\ref{fig:not_matched_sci} shows the distribution of \texttt{Science-Quality} flares that fail to match an event in either the operational \texttt{SWPC-FTP} or \texttt{SSW} catalogs. Only 58 flares of class M or higher (2 X-flares) are unmatched, and the overall discrepancy decreases substantially after 2019-12-07, when the GOES-R satellites became primarily operational.

\subsection{Science-Quality Flare List Augmentation}

We augment the \texttt{Science-Quality} flare list by attaching NOAA Active Region (AR) numbers to individual flares. Location information is essential for flare prediction, as flares arise from localized magnetic activity, and many flare-forecasting studies rely on AR identifiers to link flares to HARP patches and utilize region-specific physical features. However, the NCEI \texttt{Science-Quality} flare summary product does not include NOAA AR numbers. It only provides precise pinpoint flare locations for the GOES-R series (GOES-16/17/18/19) beginning on 2017-02-09, expressed in four coordinate systems: Stonyhurst/heliographic (longitude, latitude), Carrington coordinates, heliocentric radial coordinates, and helioprojective Cartesian (HPC) $(x,y)$~\citep{goes8-15_readme}.

For \texttt{Science-Quality} flares that are matched to a \texttt{SWPC-FTP} or \texttt{SSW} flare, we directly adopt the AR number from the matched operational report. However, approximately 37\% of science-quality flares cannot be matched, or are matched to entries associated with an invalid AR number (0). To extend AR coverage for these cases, we use the pinpoint flare locations provided by the GOES-R series, together with NOAA Solar Region Summary (SRS) reports. The SWPC SRS reports, available daily through the public FTP archive, provide detailed information for each numbered active region, including its position in Stonyhurst/heliographic coordinates. 

For every \texttt{Science-Quality} flare with valid HPC coordinates, we assign an AR number by identifying the nearest active region in the SRS report using the Euclidean distance in HPC coordinates: 
\[
\text{Distance}\;=\; \sqrt{(X_{\text{AR}}-X_{\text{Flare}})^2 + (Y_{\text{AR}}-Y_{\text{Flare}})^2}.
\]
Among the flares occurring during the interval for which precise location information is available, 93\% (11,783 events) can be assigned to a nearest AR. Table~\ref{tab:nearest_ar_distance} summarizes the Euclidean distance (in arcseconds) between a flare and the nearest AR. 

\begin{table}[!ht]
\centering
\caption{Summary statistics for Euclidean distance (in arcsec) between \texttt{Science-Quality} flare and the nearest Active Region in the helioprojective Cartesian coordinate. }
\label{tab:nearest_ar_distance}
\begin{tabular}{lcccccc}
\toprule
Count& mean& std.& min& 50\%& 75\%& 90\%\\
\midrule
6592& 151.42& 196.13& 1.50& 102.64& 174.01& 290\\
\bottomrule
\end{tabular}
\end{table}

To ensure reliable AR assignments, we retain only matches with a distance smaller than 250 arcsec. This threshold is mainly motivated by the empirical distribution of flare–AR distances (Table~\ref{tab:nearest_ar_distance}), where it lies between the 75th and 90th percentiles. As such, it effectively removes large-distance matches that are likely to correspond to incorrect associations, while preserving the majority of physically plausible matches. The threshold is also consistent with the overall scale of the distribution, being close to the mean plus one standard deviation (151.42 + 196.13 arcsec). We then assign the nearest AR number to flares that are either unmatched in \texttt{SWPC-FTP} or reported with AR number 0. As a result, 87\% of GOES-R flares have valid AR labels, compared with only 77\% before augmentation, representing an increase of approximately 10\% of all flares in this period.

{We emphasize that this nearest-neighbor assignment in HPC coordinates is primarily a pragmatic approximation and does not strictly correspond to physical surface distances, particularly for flares and active regions located away from the solar disk center. For this reason, the procedure is applied only to flare events that do not have a matched operational flare with a valid AR number. The validity of this approach is supported by an empirical consistency check. Among flares that also have a nonzero AR number reported in the \texttt{SWPC-FTP} catalog, 70\% of the nearest-neighbor assignments agree exactly with the reported AR number, and more than 95\% differ by at most 6 in AR number. This indicates that, despite its simplicity, the nearest-neighbor approach provides a reasonable approximation for AR association in the absence of direct labels.}

\begin{figure}[!htp]
    \includegraphics[width=1.1\textwidth, left]{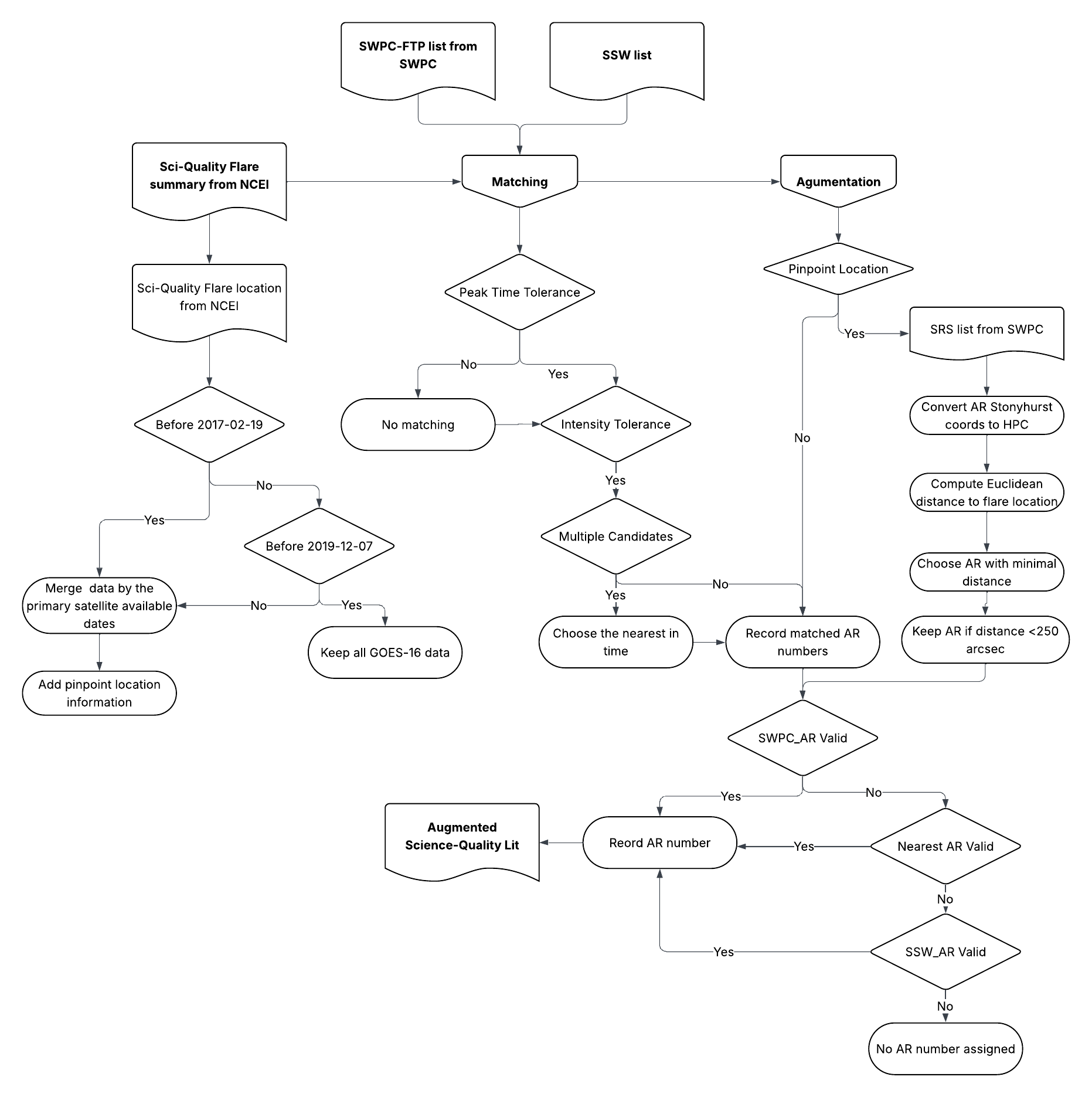}
    \caption{A diagram illustrating the data processing pipeline.} 
    \label{fig:diagram_data}
\end{figure}

For flares occurring before the availability of GOES-R pinpoint locations, AR numbers are assigned only when a valid match exists in either \texttt{SWPC-FTP} or \texttt{SSW}. Eventually, we were able to assign valid AR numbers to 72\% of all \texttt{Science-Quality} flares. This augmented science-quality flare list is used in the model-training comparisons presented in Section~\ref{sect:pred_comparison}. {A summary of the data processing steps is provided below, and the complete pipeline is illustrated in Figure~\ref{fig:diagram_data}.}

\noindent\textbf{(1) Construction of the \texttt{Science-Quality} flare list.}

We first obtain the science-quality flare summary and flare location products from the NCEI archive. Because NCEI provides these products separately for each GOES instrument, we combine data from multiple satellites to cover the full study period (2010-01-01 to 2024-12-31). Specifically, prior to the availability of GOES-R flare location products (before 2017-02-19), we merge flare records from GOES-13, -14, and -15 based on the periods during which each satellite serves as the operational primary instrument.

After the introduction of GOES-R, flare location information becomes available. For flares observed by GOES-16, we augment the flare list by linking each event to its corresponding pinpoint location using the unique identifier. We retain GOES-16 data exclusively to ensure consistency in flare location information.

\noindent\textbf{(2) Construction of comparison flare lists.}

We obtain the operational flare list from the SWPC FTP archive (\texttt{SWPC-FTP}) and the SolarSoft Latest Events (\texttt{SSW}) flare catalog from the corresponding online repository. These datasets serve as reference catalogs for the NOAA Active Region number information. 

\noindent\textbf{(3) Flare matching across catalogs.}

Flares are matched across the \texttt{Science-Quality}, \texttt{SWPC-FTP}, and \texttt{SSW} catalogs based on peak time and intensity criteria. This step provides corresponding events and AR information across catalogs.

\noindent\textbf{(4) Nearest active region assignment.}
For each \texttt{Science-Quality} flare with valid HPC coordinates, we identify the nearest active region listed in the SRS report by minimizing the Euclidean distance in HPC coordinates. The AR label is assigned from the nearest region and retained only if the distance is less than 250 arcsec.

\noindent\textbf{(5) Augmentation of AR information.}
AR labels for \texttt{Science-Quality} flares are assigned using a hierarchical procedure that depends on data availability.

For flares observed prior to the GOES-R era (before 2017-02-19), AR labels are inherited from matched events in the comparison catalogs. Specifically, if a flare is matched to a \texttt{SWPC-FTP} event with a nonzero AR number, that AR label is adopted. Otherwise, if a match is found in the \texttt{SSW} catalog with a nonzero AR number, the corresponding AR label is used. 

For flares observed in the GOES-R era (after 2017-02-19), we first assign AR labels using the matching-based procedure described above. For flares that cannot be matched to either catalog, AR labels are assigned based on the nearest active region, as described in Step (4).

\section{Flare Predictor: Data Source, Quality, and Processing}\label{sect:pred_quality}


\subsection{Imaging Predictors: HMI/SDO, AIA/SDO, and SOHO/MDI}\label{subsect:image_preds}


The imaging dataset of interest is the imaging data provided by the Atmospheric Imaging Assembly (AIA) \citep{lemen2012atmospheric} and Helioseismic and Magnetic Imager (HMI) \citep{scherrer2012helioseismic} instruments from the Solar Dynamics Observatory (SDO) since its launch in 2010 \citep{pesnell2012solar}. These instruments capture the full-disk images of the Sun in high spatial resolution (4096 $\times 4096$) with high temporal frequency as well. However, the data from HMI and AIA do not align up one-to-one as the AIA images have a larger field of view, resulting in different pixel sizes.  There has been extensive work on processing SDO instrument data to facilitate its use in machine learning research, given these discrepancies. For example, \citet{galvez2019machine} curates a data set by removing poor quality images, rescaling the images to a lower and more manageable resolution (512 $\times 512$), and applying corrections due to machine degradation and exposure issues. 

In particular, a data product mentioned in \citet{chen2024solar} processes the existing HMI HARP data with the AIA images (in the 8 channels of 94, 131, 171, 193, 211, 304, 35, and 1600 Angstroms, {denoted by $\AA$ hereafter}) to make it the same field of view and coordinates, which is tailored specifically for machine-learning-based flare prediction research. This data set contains all B, C, M, and X-class flares from 2010-2024, starting 24 hours before the flare peak-time at a temporal cadence of 12 minutes. The flare peak time is reported by the Geostationary Operational Environmental Satellite (GOES), operated by the United States' National Oceanic and Atmospheric Administration (NOAA). The HMI instrument is used to study oscillations and the magnetic field of the Sun's photosphere. The image contains 2D photospheric maps of the 3 orthogonal magnetic components. This contrasts with the MDI instrument on SOHO, which recorded only 2D line-of-sight magnetic-field images; HMI contains the radial field component of the magnetic image, $B_r$. This orthogonal component of the magnetic field is not only where some of the summary statistics (like the SHARPS that we will discuss further in Section \ref {subsect:SHARPs_SMARPs}) for the Sun are derived from, but also where the magnetic polarity inversion lines (PIL) can be identified from. This is important because the PIL is known to be a key driver of flares and other space weather phenomena \citep{sun2023tensor}.

\begin{figure}[!htb]
    \centering
    \includegraphics[width=\textwidth]{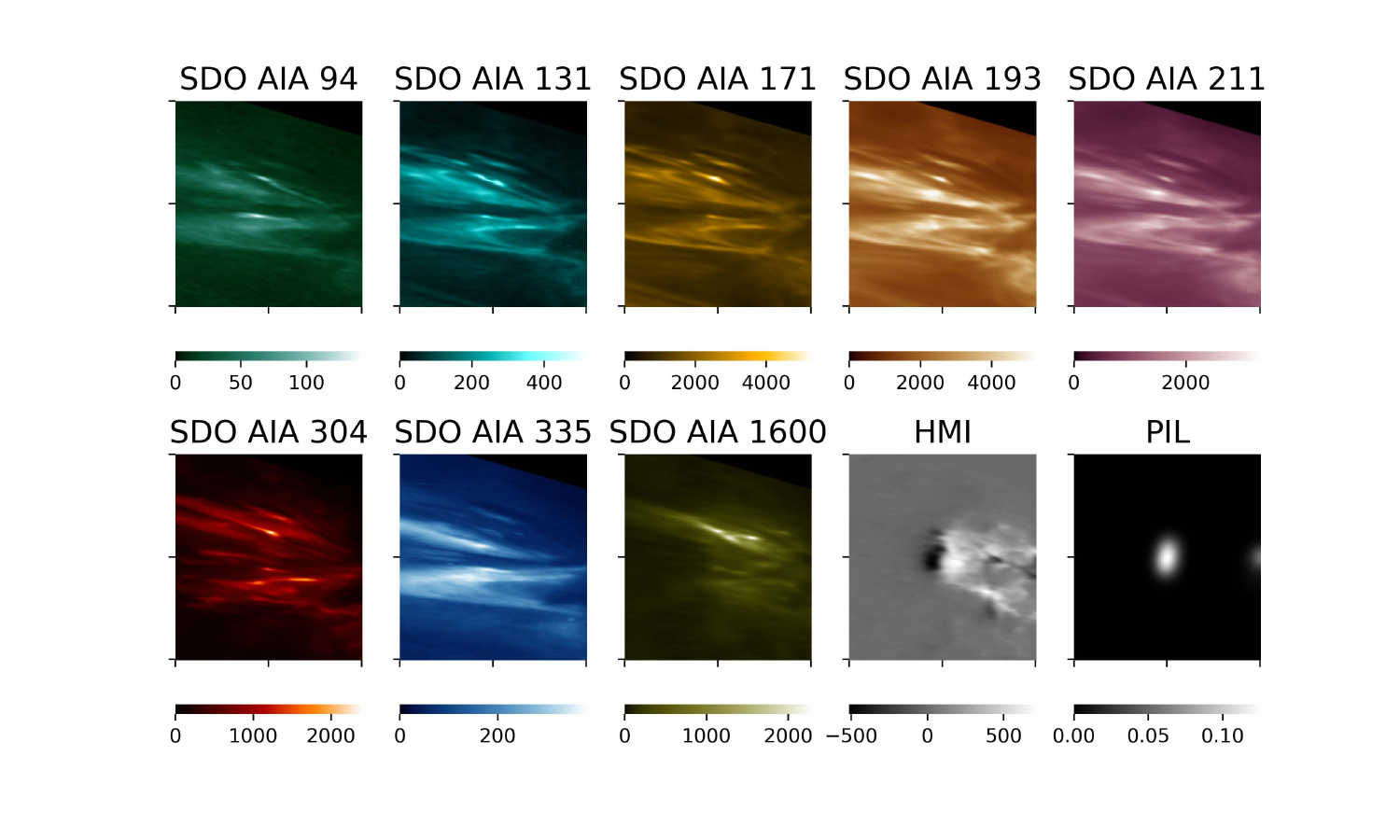}
    \caption{AR11079 Example Image at 2010-06-13T023100 which is on the limb with a max longitude that is around 80 degrees. It is off-center, blurry, and difficult to discern specific details. 
}
    \label{fig:on_limb}
\end{figure}

\begin{figure}[!htb]
    \centering
    \includegraphics[width=\textwidth]{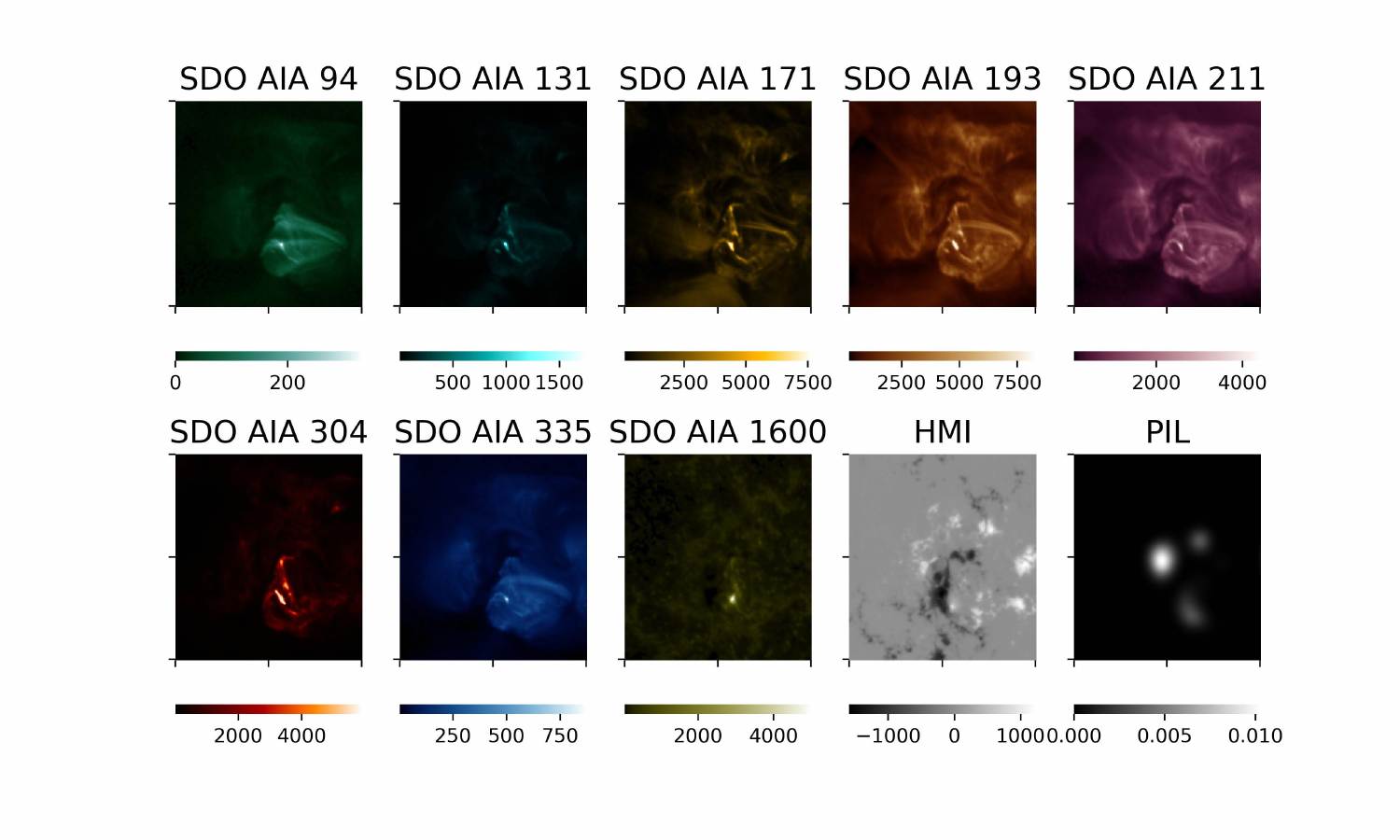}
    \caption{AR11121 Example Image at 2010-11-11 04:36:00 which is a clear, centered image not on the limbs of the instrument. 
}
    \label{fig:off_limb}
\end{figure}

One issue with the solar imaging data from the HMI and AIA concerns ``near-limb'' images, those images of active regions at the limb, which can introduce potential data-quality issues. Table \ref{tab:hmi_missing_limb} shows the mean proportion of missingness for images near the limb and in the Disk center (which we define to have a longitude of above 68 degrees or less than -68 degrees, like in \citet{chen2019iden}). As shown, the proportion of missing data is significantly higher when the images are near the limb, with greater variability. Figure \ref{fig:on_limb} is an example near-limb image of AR11079. There are clear quality issues with this image: not only is it blurry and skewed, but the specific details and contrast are not very distinct. In contrast, Figure \ref{fig:off_limb} from AR11121 shows a nearly perfectly centered AR. The quality difference is substantial, especially given that the specific details are readily discernible. As such, when dealing with these solar flare images, it is best to focus on images that are not on the limb for quality control. After such a procedure, we can curate a list of images for use in modeling or other data tasks.

\begin{table}[ht]
    \centering
    \begin{tabular}{|c|r|r|r|} \hline 
         Region & Mean Prop Missing & Std. Dev. & Total Image Counts \\
         \hline
         Near Limb  & 0.114 & 0.188 & 424,627 \\
         Disk Center & 0.00014 & 0.0035 & 522,051 \\
         \hline
    \end{tabular}
    \caption{Mean proportion of missingness for images on the limb versus disk center along with total image counts}
    \label{tab:hmi_missing_limb}
\end{table}


Another concern with the solar flare imaging data is the sporadic missingness in some ARs. Ideally, a continuous 24-hour sequence of an active region’s evolution prior to flare onset would be used as the predictors for machine learning models. In practice, such complete coverage is uncommon because the active region must be identified early enough and remain in view for the entire 24 hours. In addition, AIA data gaps can occur, meaning that even for regions currently being observed, some time intervals are not recorded, creating gaps in otherwise continuous video data. Due to the cyclic nature of the Sun's rotation as well. We will have gaps in observations of active region patches on the Sun because, when they rotate out of our view, we will no longer see them until they rotate back. This means that if there were any solar activity, such as the formation of a new active region or the end of one, the specific time and moment may not be captured. 

\begin{figure}[ht]
    \centering
    \includegraphics[width=1\textwidth]{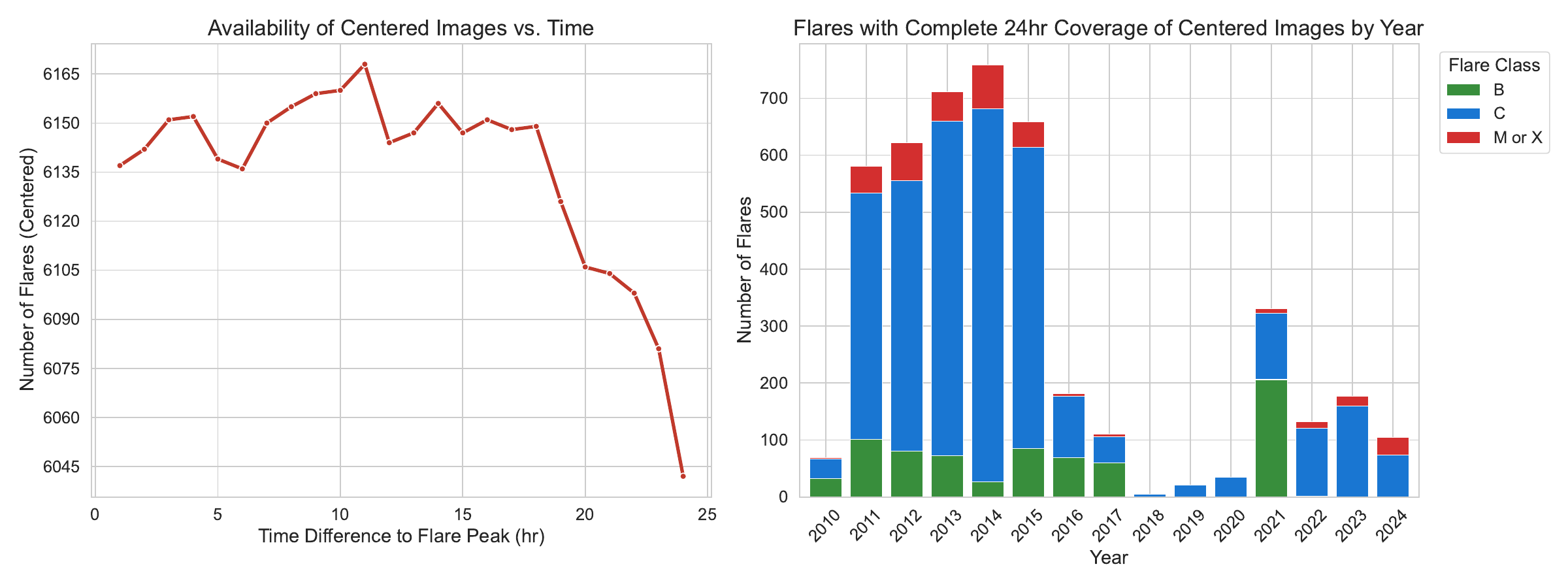}
    \caption{
    Temporal coverage of on-disk flare images. (Left) Count of unique flares with available data not on the limbs as a function of the nearest hour until flare onset. (Right) Stacked annual counts of unique flares that possess complete (non-limb images) 24-hour data coverage, classified by X-ray intensity (M/X, C, and B classes).
    }
    \label{fig:flare_img_prop_hr}
\end{figure}

For the prediction of solar flares, it is desirable to use historical imaging data of active regions 24 hours before an eruption as training data. However, if we use a flare list listing all flares from 2010-2024 and their associated active regions, and examine images available from 24 hours before peak time to peak time, we find that there is generally less data available the further we attempt to forecast. This visualization is in Figure \ref{fig:flare_img_prop_hr}.
    
\subsection{Vector Predictors: SHARPs}\label{subsect:SHARPs_SMARPs} 

If a machine learning model uses images for prediction, it must be trained on a large volume of images to perform well. As stated in Section~\ref{subsect:image_preds}, HMI images are high-resolution. High-resolution images generate large files and are thus challenging to store in large quantities and to process efficiently within training pipelines. An alternative to using images is to use vectors of summary statistics computed from images; the SHARP dataset is based on this idea. The following description of the dataset draws primarily on \cite{SHARPs}.

SHARP stands for Space-weather HMI Active Region Patch. An HMI Active Region Patch, or HARP, is a magnetic structure akin to an active region \citep{HARPs}, though a HARP can contain multiple NOAA active regions \citep{SHARPs}. A software pipeline automatically detects HARPs in HMI images and tracks them as they traverse the solar disk \citep{HARPs}. The pixels within a HARP are used to calculate the vector of sixteen SHARP parameters, each of which describes some aspect of the vector magnetic field of the active region within the HARP. See Table 3 in \citet{SHARPs} for the definitions and explanations of the SHARPs. For example, the parameter \texttt{USFLUX} measures the total unsigned flux, and the parameter \texttt{MEANJZD} measures the vertical current density. Previous studies have demonstrated that the SHARP parameters exhibit predictive power for flare forecasting \citep[e.g.,][]{bobra2015sola, chen2019iden}, supporting the notion that summary statistics can be an effective alternative to images.

There are two types of HARP products: definitive (DEF-SHARPs) and near-real-time (NRT-SHARPs). \footnote{See \url{http://jsoc.stanford.edu/HMI/HARPS.html} for details.} NRT-SHARPs are generated within a few hours of data acquisition and therefore provide more timely access, but their parameter quality is generally lower. Definitive SHARPs become available approximately five weeks later and incorporate improved calibration and processing. It is important to note that HARP identifiers are not consistent between the definitive and NRT products. The DEF-SHARPs dataset spans two additional years beyond the availability of NRT-SHARPs (starting on 2010-05-01 rather than 2012-09-14), yielding approximately 2.3 million additional measurement samples not present in the NRT set. Even within the overlapping period, definitive SHARPs include 912 more unique NOAA active regions (ARs) than the NRT product, with only 1043 ARs present in both datasets. 
We compare these two products in terms of prediction model performance in Section~\ref{sect: def_vs_nrt}.

Many other quantities are computed alongside the SHARP parameters, including various longitudes and latitudes. The bits of the \texttt{QUALITY} keyword indicate whether various issues, like eclipses, make the calculated values unreliable\footnote{See \url{http://jsoc.stanford.edu/jsocwiki/Lev1qualBits} for all the issues that can be flagged.}. Even if the \texttt{QUALITY} keyword does not indicate any problems, some SHARP parameters can still be missing; at other times, the entire record, i.e., the SHARP parameters, longitudes and latitudes, \texttt{QUALITY}, etc., may be missing. Keywords that can only be computed when a magnetic field is present can have missing values during ``padding intervals''. Keywords can also have missing values for HARPs that are faint \citep{HARPs}. We do not know if these are the only possible causes of missingness.


We investigated data quality issues in the HARP and SHARP data for the period 1 May 2010 to 21 August 2024. The former date is the earliest date for which SHARP data is available \citep{SHARPs}. We excluded HARP records for near-limb HARPs as the data for those HARPs is too corrupted by noise to be usable; we classified records for which the Stonyhurst longitude of the flux-weighted center (\texttt{LON\_FWT}) was beyond $\pm68^{\circ}$ as near-limb, based on \cite{bobra2015sola}. For any HARP, the times for consecutive SHARP parameter vectors differ by twelve minutes. The time of a record can be the start of an hour, 12 minutes after the start, 24 minutes after the start, etc. For every HARP, if a time that should have had a record for the HARP did not, we created a record with missing values for the SHARP parameters and other variables. At any time whose minutes value is divisible by 12, there may be one or more HARPs in existence, or there may be no HARPs in existence. In the former case, at the given time, at least one SHARP vector would be available, and in the latter, no SHARP vectors would be available. For each time in the specified period, we determined whether any SHARP vectors were available for it. We found that 566,100 (90\%) of the times had at least one SHARP vector, and 61,140 (10\%) did not have any SHARP vectors.

\begin{figure}[!htb]
    \centering
    \includegraphics[width=\textwidth]{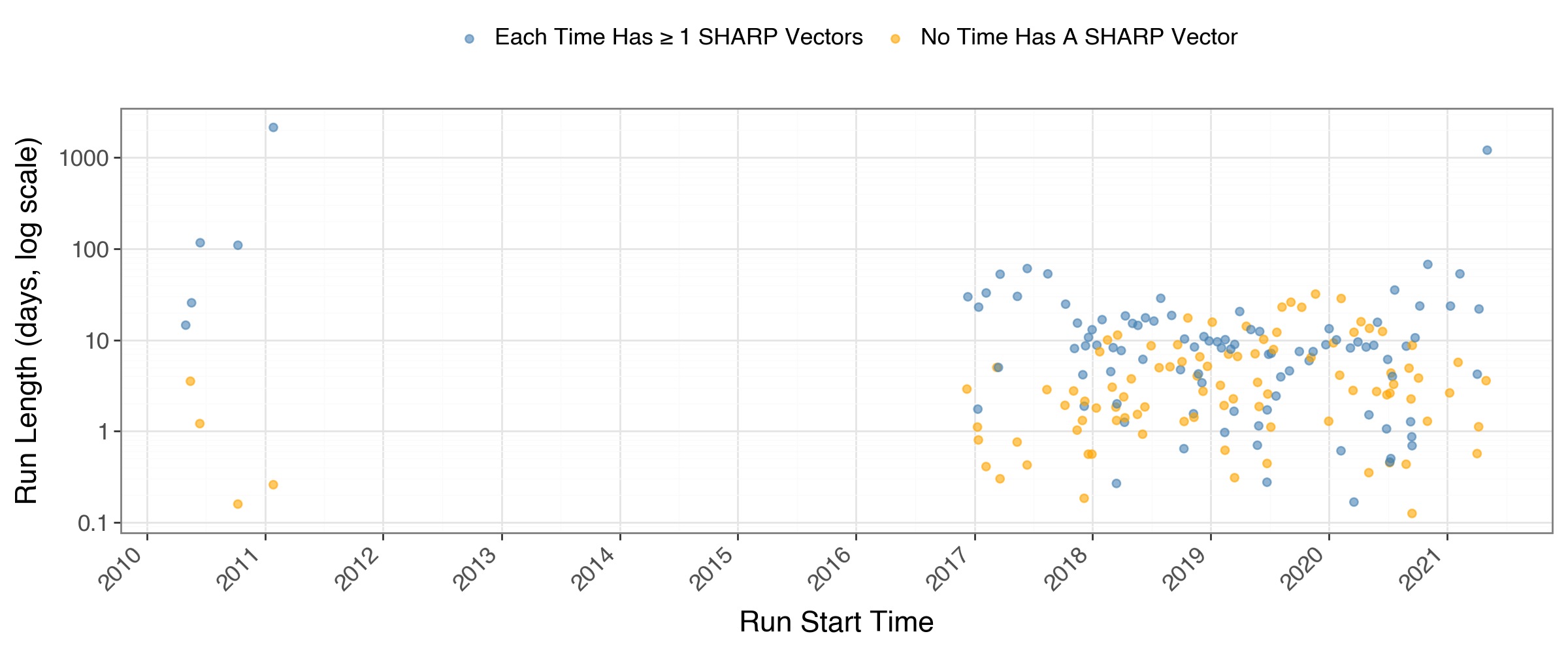}
    \caption{Each dot in this plot represents one run of consecutive times in the sequence of times between 1 May 2010 and 21 August 2024 that could be the time of a SHARP vector, i.e., has a minutes value that is divisible by 12. A dot is colored blue if each time in the corresponding run has at least one SHARP vector and orange if no time in the run has a SHARP vector. The first coordinate of the dot gives the start time of the run (the earliest time in the run), while the second gives the run length (last time minus earliest time) in days.}
    \label{fig:run_length_vs_start_time}
\end{figure}

The sequence of times can be partitioned into runs such that, within each run, either every time has at least one SHARP vector or no time has a SHARP vector. For example, suppose that whether each time in the left column of Table~\ref{tab:runs_example} had at least one SHARP vector was given by the middle column (the values in the middle column were made up for this example).
\begin{table}[]
    \centering
    \begin{tabular}{|c|c|c|}
        \toprule
        Time & Has $\ge$ 1 SHARP Vector? & Run ID \\
        \toprule
        1 May 2010 00:00 & Yes & 0 \\
        \midrule
        1 May 2010 00:12 & Yes & 0 \\
        \midrule
        1 May 2010 00:24 & No & 1 \\
        \midrule
        1 May 2010 00:36 & No & 1 \\
        \midrule
        1 May 2010 00:48 & Yes & 2 \\
        \bottomrule
    \end{tabular}
    \caption{An example illustrating the concepts of a run of times that have at least one SHARP vector and a run of times that do not have any SHARP vectors. The times in the first two rows form a run of the former type, and the times in the third and fourth rows form a run of the latter type. The values in the middle column were made up for this example.}
    \label{tab:runs_example}
\end{table}
Then 1 May 2010 00:00 and 1 May 2010 00:12 would form a run in which each time has at least one SHARP vector, and 1 May 2010 00:24 and 1 May 2010 00:36 would form a run in which no time has a SHARP vector. For each run, we computed its start time (the earliest time in the run), its end time (the last time in the run), and its duration (the difference between the start and end times). These runs are visualized in Figure~\ref{fig:run_length_vs_start_time}. In the figure, each dot represents one run of times. A run is blue if it consists exclusively of times with at least one SHARP vector; a run is orange if it only contains times that do not have a SHARP vector. The first coordinate of a dot is the start time of the corresponding run, while the second coordinate is the run’s duration in days. Orange runs should not be used, as there is no HARP data for them and no way to impute it. It seems reasonable to train a model with at least several months’ worth of data; there are only two blue runs that could be divided into training sets with that much data, the two runs whose lengths are over 1,000 days. Both of those runs correspond to solar maxima. Table~\ref{tab:longest_good_runs} displays the five longest runs that have at least one SHARP vector at each time. They correspond to the five blue dots in Figure~\ref{fig:run_length_vs_start_time} with the highest run lengths. As stated above, only two runs are long enough that they can be carved into multiple reasonably-sized training sets; those two runs are in the first two rows of Table~\ref{tab:longest_good_runs}. In the rest of this section, we will only consider times from the longest run, 22:00 on 26 January 2011 to 14:48 on 8 December 2016.

\begin{table}[!htb]
    \centering
    \begin{tabular}{|c|c|c|c|}
        \toprule
        Run Start Time & Run End Time & Run Length (Days) & Run Length (Times) \\
        \toprule
        2011-01-26 22:00 & 2016-12-08 14:48 & 2,142.71 & 257,125 \\
        \midrule
        2021-05-04 05:12 & 2024-08-21 23:48 & 1,205.78 & 144,694 \\
        \midrule
        2010-06-14 19:12 & 2010-10-09 04:24 & 116.39 & 13,967 \\
        \midrule
        2010-10-09 08:24 & 2011-01-26 15:36 & 109.31 & 13,117 \\
        \midrule
        2020-11-01 15:48 & 2021-01-08 00:12 & 67.36 & 8,083 \\
        \bottomrule
    \end{tabular}
    \caption{The sequence of times between 1 May 2010 and 21 August 2024 was divided into runs of consecutive times in which either each time has a SHARP vector or no time has a SHARP vector. This table shows the start and end times of the five longest runs of the former type. A machine learning model cannot be trained or used to generate predictions during a run of the latter type. Lengths are given both in days and in times.}
    \label{tab:longest_good_runs}
\end{table}

A HARP record can have one or more problems. A record can have missing values for all SHARP parameters; we call such records missing. A record can be missing values for some, but not all SHARP parameters; we call such records incomplete. Some records are marked as low-quality (\texttt{QUALITY} has a nonzero value). We flagged records with these problems. As Table~\ref{tab:problem_breakdown} shows, most records do not have any of these problems, but a sizable fraction have some problem. Well under 1\% of records have more than one problem. The most important subsets of problematic records are those that are only low-quality (9\% of records) and those that are only missing (8\% of records).

\begin{table}[!htb]
    \centering
    \begin{tabular}{|p{2cm}|p{2cm}|p{2cm}|p{2.1cm}|p{2cm}|p{2.1cm}|}
        \toprule
        Is Record \newline Missing? & Is Record \newline Incomplete? & Is Record \newline Low-Quality? & Number of \newline Records & Percentage of \newline Records \\
        \toprule
        No & No & No & 3,466,955 & 83\% \\
        \midrule
        No & No & Yes & 353,867 & 9\% \\
        \midrule
        Yes & No & N/A & 334,403 & 8\% \\
        \midrule
        No & Yes & No & 4,615 & $\approx$ 0\% \\
        \midrule
        No & Yes & Yes & 440 & $\approx$ 0\% \\
        \bottomrule
    \end{tabular}
    \caption{A breakdown of the SHARP dataset by data quality problem. A record is missing if the values for all SHARP parameters are missing. A record is incomplete if it has missing values for some SHARP parameters, but not all. A record is low-quality if the \texttt{QUALITY} keyword is nonzero, which indicates that the record has a quality issue. 83\% of records have no problems, and less than 1\% have multiple problems.}
    \label{tab:problem_breakdown}
\end{table}

\section{Comparison of Data Products from a Flare Prediction Perspective}\label{sect:pred_comparison}

We compare the \texttt{SWPC-FTP} operational flare list with the NCEI \texttt{Science-Quality} list by evaluating the numerical skill scores of machine learning models trained on definitive or near-real-time SHARP summary vector parameters. Note that we do not exhaust the available machine learning models and do not aim to optimize reported skill scores here, since, as discussed in the Introduction, (i) when properly trained, commonly used machine learning models do not generate significantly different results, and (ii) the skill scores of machine learning models depend heavily on sample selection and sample split schemes. Therefore, we follow standard practice in selecting the machine learning model and sample split to demonstrate the effects of the various data products on the trained model's performance.


\subsection{Data Preparation and Model Setting}

To assess how data quality affects flare forecasting performance, we consider two sources of flare labels (the \texttt{SWPC-FTP} list and the \texttt{Science-Quality} list) and two versions of predictor data (HMI near-real-time SHARPs and definitive calibrated SHARPs), as shown in Table~\ref{tab:combo}. We evaluate two representative model families: a deep learning architecture (LSTM) and a conventional statistical model (logistic regression with PCA, i.e., principal component analysis). All models are trained on Solar Cycle 24 (2010-05-01 to 2019-12-31) and tested independently on years 2020–2024. The task is formulated as the prediction of $\geq$ M-class flares in the next {24, 12, 6} hours.

{We denote each model configuration using the format Model–FlareList–-SHARPType–Window. For example, LSTM-S-DEF-24 refers to an LSTM model trained using the \texttt{Science-Quality} flare list (S), with definitive SHARP inputs (DEF), and a 24-hour forecasting window. Similarly, ``Logreg'' denotes the Logistic regression model, ``O'' denotes the \texttt{SWPC-FTP} (operational) flare list, and ``NRT''  denotes near-real-time SHARP data. This naming convention is used consistently throughout the figures and tables to distinguish between different training data sources and forecasting settings.}

\begin{table}[!ht]
\centering
\renewcommand{\arraystretch}{1.3}
\begin{tabular}{l|c|c}
\hline
 & \texttt{SPWC-FTP} & \texttt{Science-Quality} \\ \hline
NRT SHARPs       & O-NRT & S-NRT \\ \hline
Definitive SHARPs & O-DEF & S-DEF \\ \hline
\end{tabular}
\caption{Combinations of flare response datasets and SHARP predictor data used in model comparison. For every combination, we train multiple LSTM-based classifiers and logistic regression models.}
\label{tab:combo}
\end{table}

We follow the data preparation and sample construction methods described in ~\citet{Jiao2020lstm}. The input data consists of 24-hour time series extracted from the full time series of SHARP parameters of ARs. To ensure data quality, some sequences are excluded, particularly when the ARs are located near the limb. The criteria for excluding unqualified time sequences are as follows. 
\begin{enumerate}
    \item To minimize projection effects, the longitude of the HARP region must fall within $\pm 70^\circ$ of the central meridian.
    \item The starting time of the two adjacent time sequences is separated by at least 1 hour.
\end{enumerate}
The response for each input sequence (predictors) is defined as the maximum flare class produced by the corresponding AR within the forecasting window of {24, 12, or 6 hours} following the end time of the input sequence. If no flare occurs within this forecasting window, the prediction window is skipped, and the time window is advanced by the stride of 1 hour. Consequently, the target labels can be categorized as ``A'', ``B'', ``C'', ``M'', and ``X''.

We focus on a binary classification problem that distinguishes strong(M/X) from Weak (A/B) flares. We exclude the C-class category {in this study}. C-class flares may be comparable to either small M-class flares or strong B-class flares, making it an unreliable separation boundary and thus vague the training process \citep {chen2019solar}. In this study, we adopt two models: an LSTM model with a sigmoid output layer and a logistic regression model. {These two are selected as two representative and well-established forecasting models with complementary properties: logistic regression provides a transparent linear baseline that is widely used in flare-forecasting studies, whereas LSTM represents a nonlinear sequence model capable of using temporal information in the input data. Moreover, previous benchmark analysis has shown that these two model produce competitive and robust performance as compared to others~\citep{Wang2020lstm,sun2022predicting}.} 

The LSTM-based model is widely used in solar flare forecasting for handling time-series inputs~\citep{chen2019iden, Liu2019lstm, Wang2020lstm} and has been shown to outperform other machine learning models~\citep{Wang2020lstm}. The network consists of a two-layer LSTM with a hidden size of 30 units and dropout of 0.3 between layers, followed by a ReLU activation and a fully connected layer that maps the final (or truncated) hidden representation to a single scalar logit. This logit value is interpreted as the pre-sigmoid score for the probability of a strong flare. We include logistic regression due to its simplicity, which in fact results in stability for trained models and interpretability. Given the $k$-hour forecasting window used to extract SHARP parameters, the input of the LSTM model for each target is a $5k\times20$ matrix, where each row represents a 20-dimensional vector of SHARP parameters. For the logistic model, we first flatten each input along the time dimension into a $100k\times1$ vector. 

The data are split into training and validation subsets using a 90/10 stratified split to preserve the positive/negative class ratio. To improve generalization and account for sampling variability, we employ a bootstrap ensemble strategy for both LSTM and logistic regression models. Specifically, we draw 30 bootstrap replicates with replacement from the training set. For each replicate, we fit the pipeline that integrates the following steps.
\begin{enumerate}
    \item Standardization.
    \item For the logistic regression model, we then apply principal component analysis (PCA) to reduce dimensionality, retaining 98\% variance. For the 24-hr forecasting window, 20 principal components already account for over 98\% of the variance. The projection coefficients for these principal components are derived from the training set.
    \item The LSTM model is optimized using the Adam's optimizer with a binary cross-entropy loss.
    \item The logistic regression model is fitted with $L_2$ regularization. The penalty strength is optimized with 5-fold cross-validation and the ROC-AUC (Receiver Operating Characteristic and  Area Under the Curve) scoring objective.
    \item To address class imbalance, we apply class-weighted loss functions in both models, where the positive class weight is set to the ratio of negative to positive samples in the training data.
    \item The class weights are computed separately for each bootstrap split.
\end{enumerate}

After each bootstrap model is trained, we evaluate probabilistic predictions on the held-out validation set and determine the optimal decision threshold that maximizes the True Skill Statistic (TSS) over a grid of candidate thresholds in $(0,1)$. The optimal threshold and corresponding validation TSS scores are recorded for every bootstrap sample. To evaluate the performance of the machine learning algorithms, we split the samples (time series of SHARP parameters and their corresponding maximum flare magnitudes) into training and test sets. Note that splitting by Active Region (AR) or HARP is crucial to prevent information leakage from the training set to the testing set. In this study, all training/testing splits are performed chronologically by HARP, which mimics the operational setting but may not result in the highest skill scores. See \citet{wang2020solar} for detailed discussions on sample splitting schemes and their impacts on flare prediction results. 

It is important to note that varying definitions of positive and negative samples make direct comparisons of results from existing literature challenging, as discussed in \citet{chen2024solar}. The purpose of this {section} is not to propose a universal method for sample construction but rather to compare model performance across different datasets. We also demonstrate numerically that differences arising from data inconsistencies or quality are unaffected by the choice of sample construction method.

\subsection{Science-Quality versus Operational ``Response''}

To quantify the effect of flare-label quality on the skill scores of the prediction models, we compare an LSTM model and a logistic regression model for 6/12/24 forecasting, trained using \texttt{Science-Quality} and \texttt{SWPC-FTP} flare lists, while using the definitive SHARPs parameter as inputs.

We trained models using data from Solar Cycle 24 (5/1/2010–12/31/2019) and tested them using different time intervals in Solar Cycle 25. The evaluation spans four distinct testing intervals that correspond to different phases of Solar Cycle 25. The 2020–2021 period reflects the early rising phase and is near solar minimum, characterized by extremely low flare occurrence and a highly imbalanced dataset. The 2022 period represents an evolving phase with increasing magnetic complexity and flare frequency. The 2023–2024 interval corresponds to the solar maximum, featuring intense flare activity and a high positive sample ratio (often exceeding 0.8). Finally, the combined 2020–2025 evaluation window averages performance across the full range of solar activity and reflects the model's overall performance with a balanced test set. 

\begin{figure}[!htb]
    \centering
    \includegraphics[width=\textwidth]{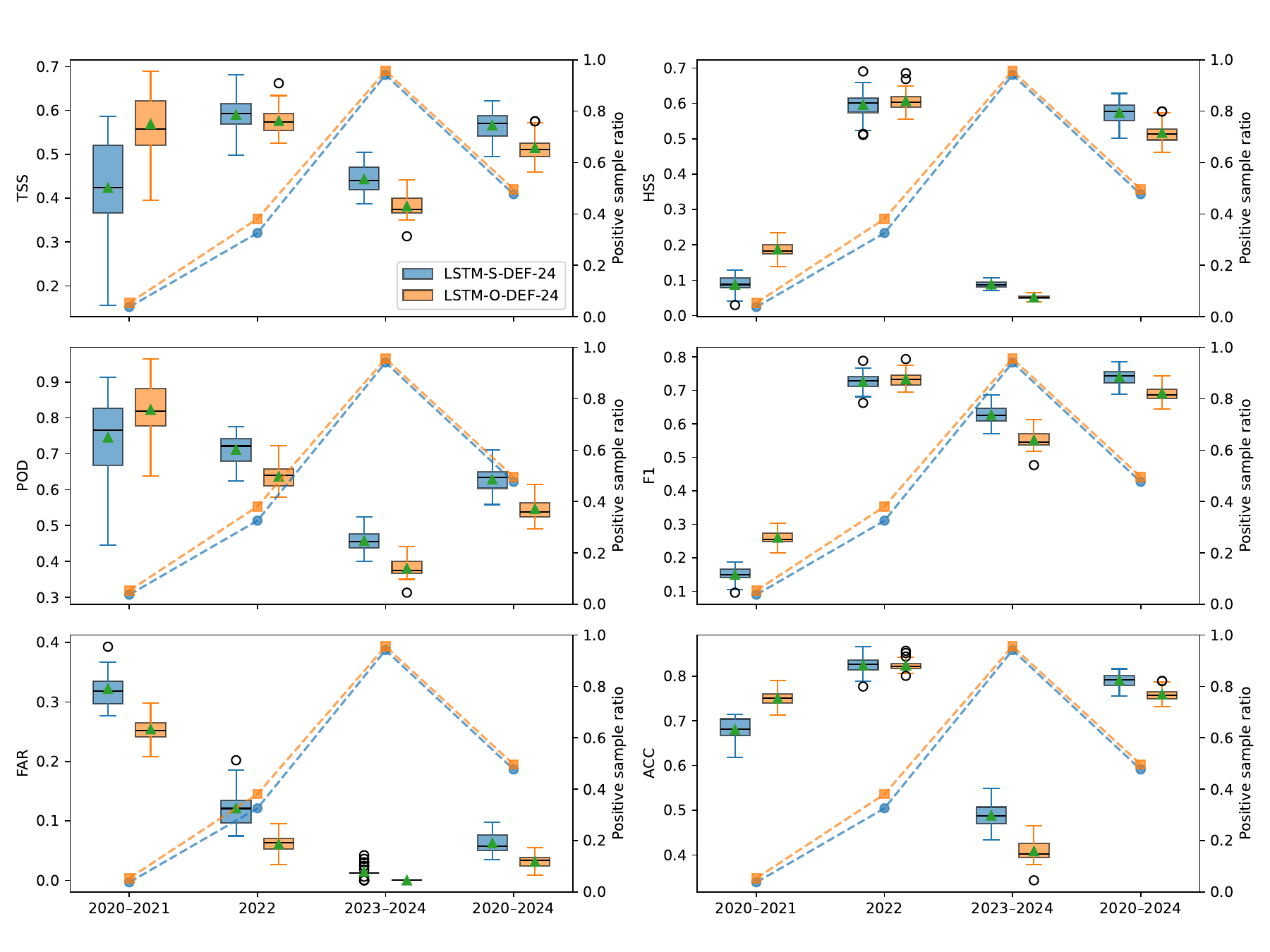}
    \caption{Box plots of six skill scores for different testing year choices (given in the x-axis) in the solar cycle 25 of the M/X versus A/B classification task by the LSTM model trained on definitive SHARPs as inputs with a 24-hour forecasting window. The blue boxes represent the \texttt{Science-Quality} flare list, while the orange boxes represent the \texttt{SWPC-FP} operational flare list. The dashed lines indicate the positive sample proportions in the test set, corresponding to the Y axis on the right-hand side of each sub-figure. Note that a smaller FAR means better performance. On each box, the black line marks the mean of the 30 independent bootstrap runs, and the triangle marks the median. The lower and upper bounds of the boxes correspond to the first and third quartiles $Q_1$ and $Q_3$. The upper and lower error bars are at $Q_3 + 1.5(Q_3-Q_1)$ and $Q_3-1.5(Q_3-Q_1)$, respectively. The small dots outside the boxes indicate data points that fall outside the error bars (outliers). The mean value and median are calculated, including the outliers.}\label{fig:LSTM_24_defSHARPS_SCI_vs_OPR}
\end{figure}

\begin{figure}[!htb]
    \centering
    \includegraphics[width=\textwidth]{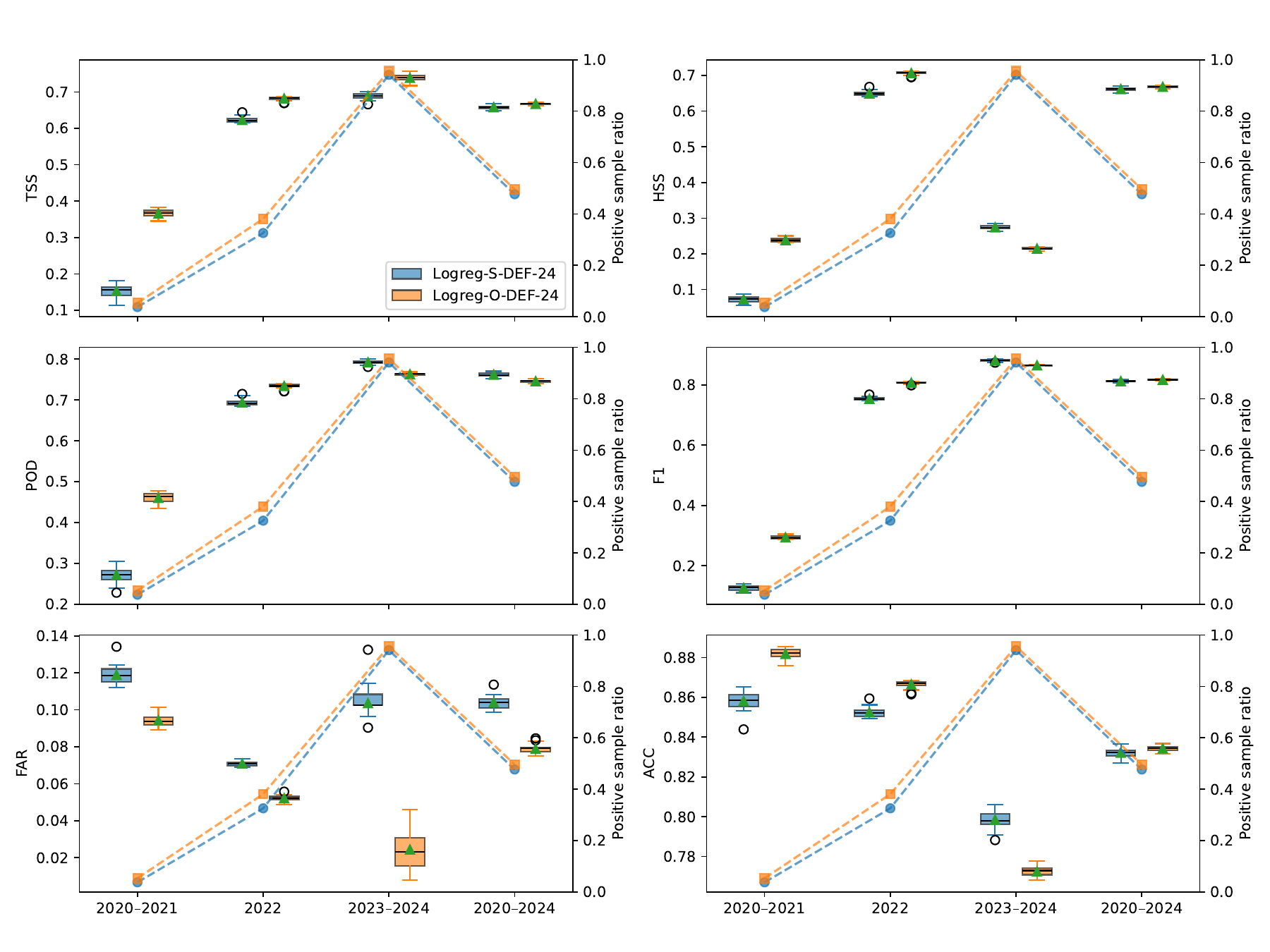}
    \caption{Same as Figure~\ref{fig:LSTM_24_defSHARPS_SCI_vs_OPR} except the model being the logistic regression.}
    \label{fig:Logreg_24_defSHARPS_SCI_vs_OPR}
\end{figure}

{We selected six skill scores (TSS, HSS, POD, F1, FAR, and ACC; see, e.g., \citet{chen2019iden} for definitions) to comprehensively evaluate model performance. TSS and HSS are widely used in space weather forecasting and are particularly suitable for imbalanced classification problems, while the remaining metrics (POD, FAR, F1, and ACC) are standard evaluation measures in general classification tasks.  Box plots of these skill scores for the 24-hour forecasting windows are shown in Figure~\ref{fig:LSTM_24_defSHARPS_SCI_vs_OPR} and Figure~\ref{fig:Logreg_24_defSHARPS_SCI_vs_OPR} for the LSTM model and the logistic regression model, respectively. The 6-hour and 12-hour window prediction results exhibit consistent trends as these. The numerical results for all model setups are presented in the Tables in~\ref{appendix:comparisonresults}.}

Across the test periods when the solar activity is relatively low, the models trained with \texttt{SWPC-FTP} operational flare list achieve higher predictive skill scores than those trained with \texttt{Science-Quality} labels. This improvement is clearer for logistic models. During the solar minimum (20200101-20211231), the logistic regression model improves from a TSS of 0.15 [0.11,0.18] (\texttt{Science-Quality}) to 0.37 [0.35,0.38] (\texttt{SWPC-FTP}), and the LSTM model improves from 0.42 [0.16,0.59] to 0.57 [0.40,0.69]. Similar margins in performance wins are observed in other metrics. For the LSTM model, the difference is not substantial. We do observe more stability with the logistic regression model results (with consistently shorter boxes in Figure~\ref{fig:Logreg_24_defSHARPS_SCI_vs_OPR} as compared to Figure~\ref{fig:LSTM_24_defSHARPS_SCI_vs_OPR}).

During the high-activity period (2023–2024), when the positive-sample ratio exceeds 0.85, the \texttt{Science-Quality}-trained models achieve better performance except for the TSS score of the logistic models. {Averaged over the entire testing period (2020–2024), the \texttt{Science-Quality}-trained LSTM models consistently outperform in all skill scores except FAR. For the logistic models, the performance differences between the two training datasets are relatively small.} Across all test periods, models trained with \texttt{SWPC-FTP} lists achieve lower FAR, demonstrating that models trained with operational flare lists are better at identifying flares without increasing false alarms.

We {then} compare the performance of different model configurations during solar minimum and solar maximum, using definitive SHARP inputs. {The results above indicate that model behavior differs substantially between these two phases of the solar cycle.} We use the logistic regression model for 6-hour forecasting trained on \texttt{SWPC-FTP} flare lists and definitive SHARPs as the baseline model. {Performance gain is quantified as the difference between the average skill score of a given model and that of the baseline (Logreg-O-DEF). We compute this gain by first averaging each skill score from bootstrapping and then taking the difference relative to the baseline. The gain can be negative by definition.} Here, we focus on the TSS and F1 for comparison {, and refer to the corresponding performance gains as TSS gain and F1 gain, respectively.}

\begin{figure}[!htb]
    \centering
    \begin{subfigure}[b]{0.48\textwidth}
        \centering
        \includegraphics[width=\textwidth]{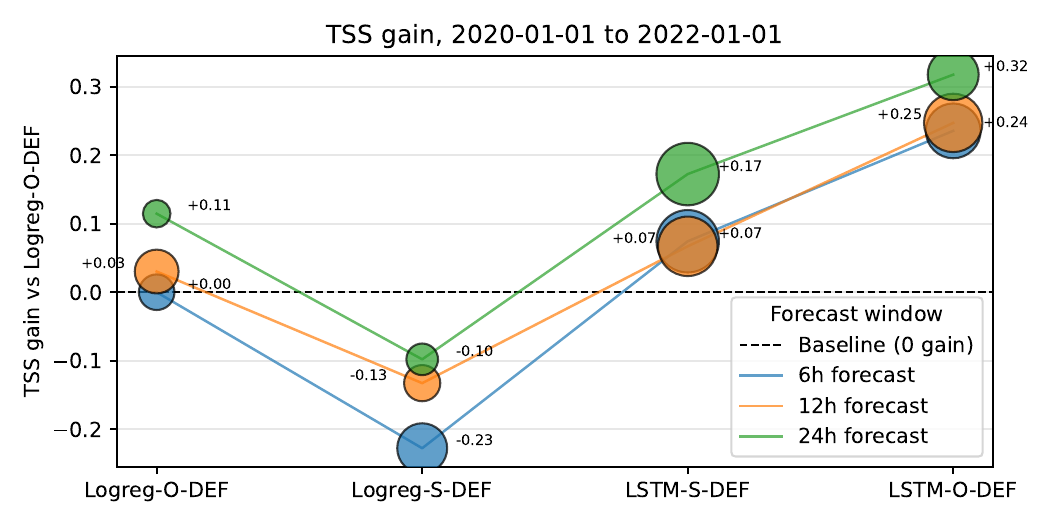}
        \caption{TSS, Solar Minimum}
    \end{subfigure}
    \begin{subfigure}[b]{0.48\textwidth}
        \centering
        \includegraphics[width=\textwidth]{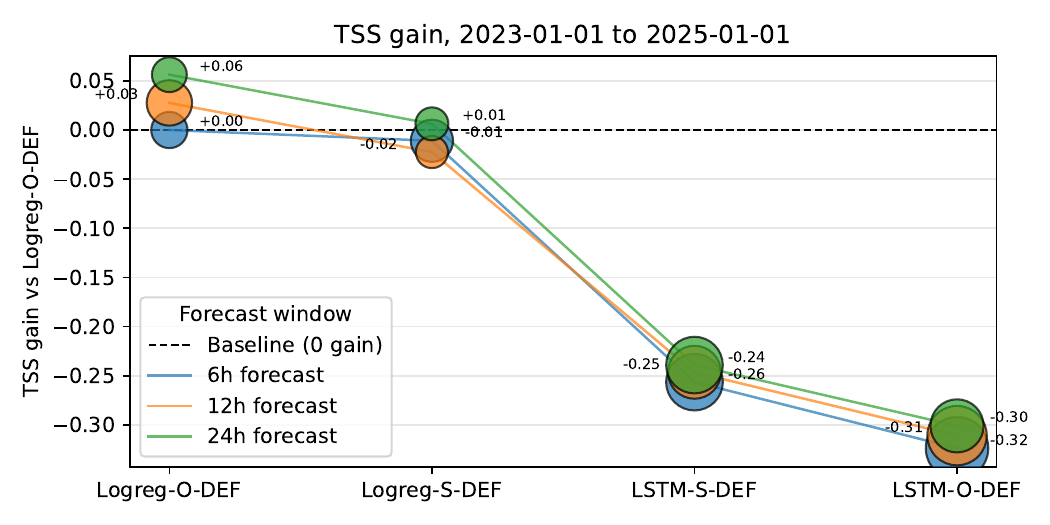}
        \caption{TSS, Solar Maximum}
    \end{subfigure}
    \begin{subfigure}[b]{0.48\textwidth}
        \centering
        \includegraphics[width=\textwidth]{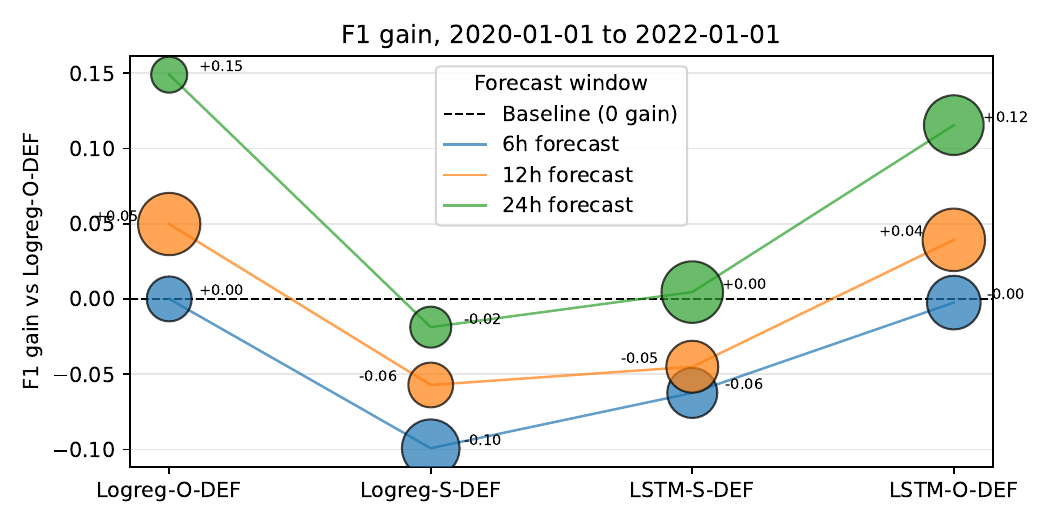}
        \caption{F1, Solar Minimum}
    \end{subfigure}
    \begin{subfigure}[b]{0.48\textwidth}
        \centering
        \includegraphics[width=\textwidth]{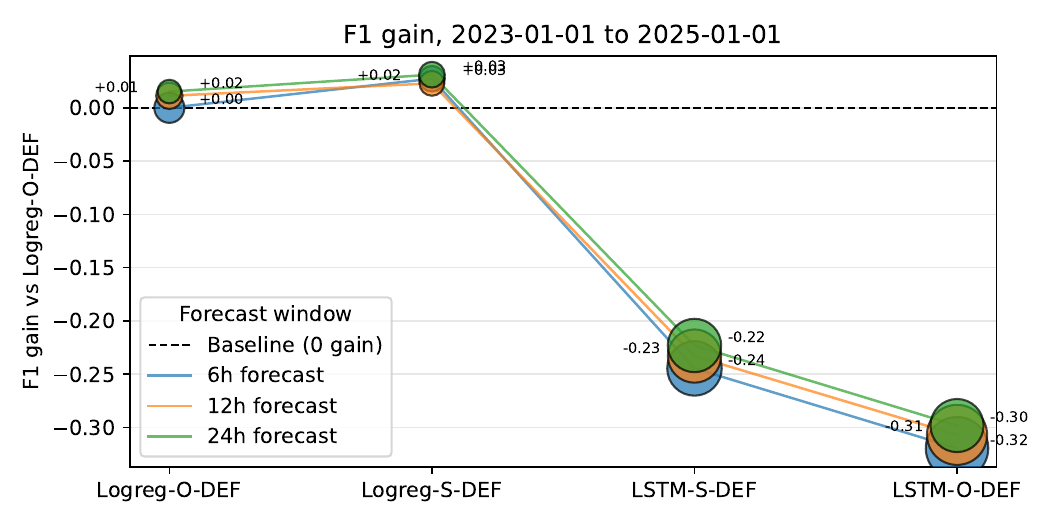}
        \caption{F1, Solar Maximum}
    \end{subfigure}
    \caption{Average performance gain relative to the baseline Logreg-O-DEF model for definitive SHARP predictors, evaluated during solar minimum (2020–2022) and solar maximum (2023–2025) testing periods. The baseline model is the logistic regression model for 6-hour forecasting trained on \texttt{SWPC-FTP} flare lists and definitive SHARPs (Logreg-O-DEF-6). {TSS gain (F1 gain) is defined as the difference between the mean TSS (F1 score) of a model and that of the baseline, averaged over 30 bootstrap realizations.} Each marker represents a model configuration (Logreg-S-DEF, Logreg-O-DEF, LSTM-S-DEF, and LSTM-O-DEF), with color indicating forecasting window length (6 h, 12 h, and 24 h). Marker size is proportional to the standard deviation across 30 bootstrap independent runs. Positive values indicate improvement over the baseline model (black dashed zero line).}
    \label{fig:metric_gain_defSHARPs}
\end{figure}

{Figure~\ref{fig:metric_gain_defSHARPs} shows the performance gain (TSS and F1 gain), relative to the baseline Logreg-O-DEF model, for different model configurations (Logreg-S-DEF, Logreg-O-DEF, LSTM-S-DEF, and LSTM-O-DEF) across forecasting windows of 6, 12, and 24 hours.} During solar minimum, LSTM models trained with the \texttt{Science-Quality} flare lists achieve higher TSS and F1 scores than the logistic regression models trained on the same labels. For the \texttt{SWPC-FTP} lists, the LSTM models perform slightly better or about the same as the logistic models. However, during solar maximum, when trained on identical flare lists, the LSTM models perform worse than the logistic models. In addition, the performance differences between LSTM and logistic models are noticeably larger during solar maximum than during solar minimum. This suggests that the more complex flare environment during solar maximum makes model choice more critical and that simpler models, such as logistic regression, can have an advantage.

\subsection{Near-real-time versus Definitive SHARP}\label{sect: def_vs_nrt}

We begin by examining the forecasting model performances when using the DEF-SHARPs and NRT-SHARPs datasets. Although the HARP numbers differ between the two products, we find that the conditional flare-response distributions (given the NOAA AR number) are highly similar. {SHARP parameters are recorded every 12 minutes, so each SHARP observation represents the state of an active region at a discrete time point. For each observation, we match flares from the \texttt{Science-Quality} list using the NOAA AR number and flare peak time. A SHARP observation is associated with a flare if the flare originates from the same NOAA AR and its peak time falls within the 12-minute interval corresponding to that observation.} Observations associated with M- or X-class flares are labeled as positive, and those associated with A- or B-class flares are labeled as negative.

\begin{figure}[!htb]
    \centering
    \includegraphics[width=\textwidth]{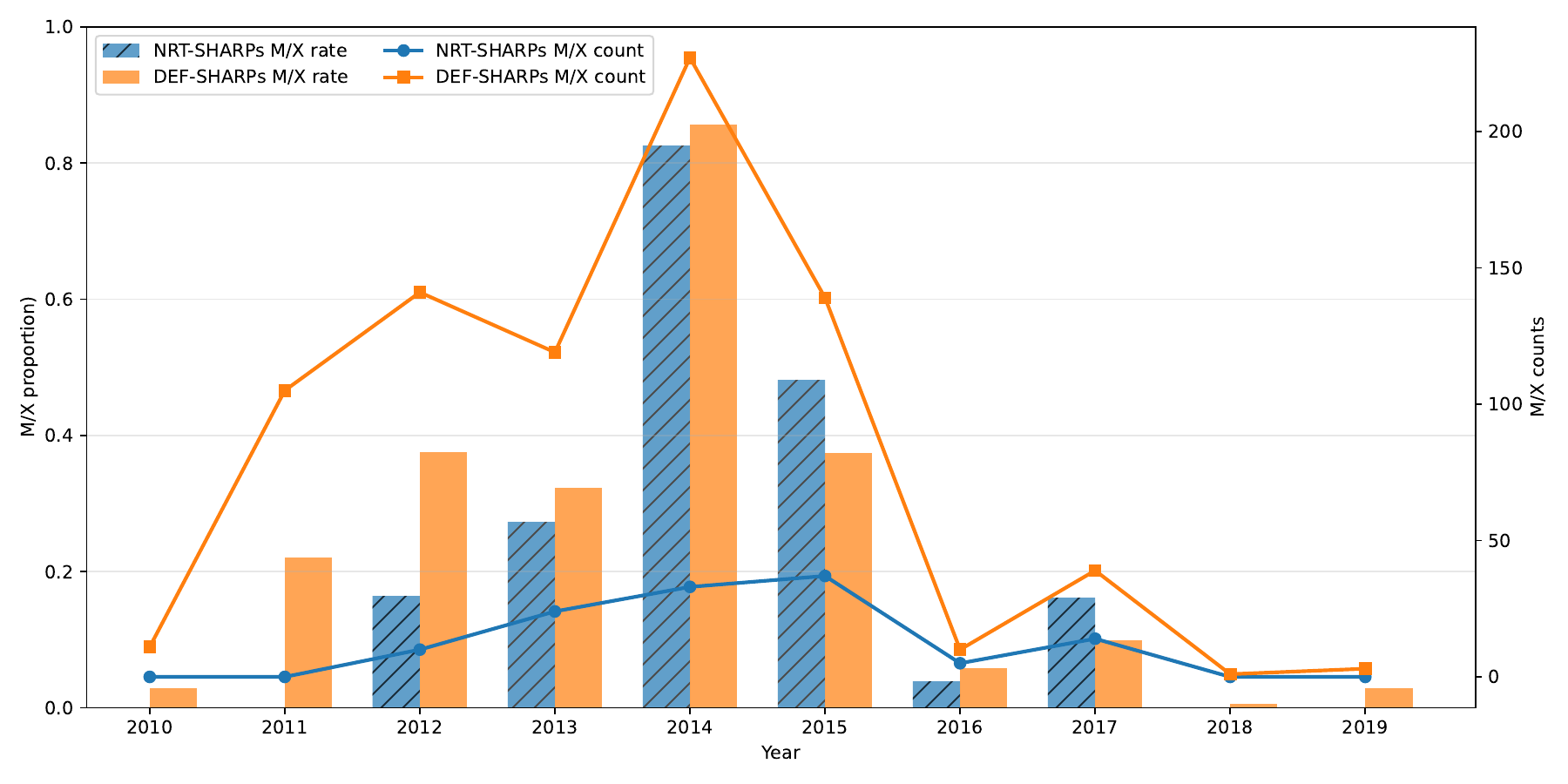}
    \caption{Comparison of yearly class imbalance in the SHARPs datasets: near-real-time (NRT-SHARPs, with blue colors) versus definitive (DEF-SHARPs, with orange colors). For each  SHARPs observation, flares from the \texttt{Science-Quality} list are matched by NOAA active-region number and flare peak time. Observations associated with M- or X-class flares are labeled as positive, while those associated with A- or B-class flares are labeled as negative. Bars show the yearly positive rate (left y-axis), i.e., the proportion of SHARPs matched to M- and X-class flares, for both NRT-SHARPs and DEF-SHARPs. The overlaid line plots (right y-axis) display the annual counts of SHARPs corresponding to positive labels (target = 1) throughout Solar Cycle 24.}
    \label{fig:def-nrt-sharp-comparison-MX-rate-and-counts}
\end{figure}

Figure~\ref{fig:def-nrt-sharp-comparison-MX-rate-and-counts} shows the yearly positive rate and positive-sample counts for both DEF-SHARPs and NRT-SHARPs. Despite fewer valid observations, the NRT-SHARPs product exhibits a positive rate pattern that closely matches that of the definitive product across all training years. {In flare forecasting tasks, flare locations are typically identified using NOAA AR numbers rather than HARP numbers. Hence, the similarity between flare-response distributions conditioned on SHARPs’ AR numbers indicates that NRT-SHARPs and DEF-SHARPs provide largely consistent geographic information.}

\begin{figure}[!htb]
    \centering
    \begin{subfigure}[b]{0.48\textwidth}
        \centering
        \includegraphics[width=\textwidth]{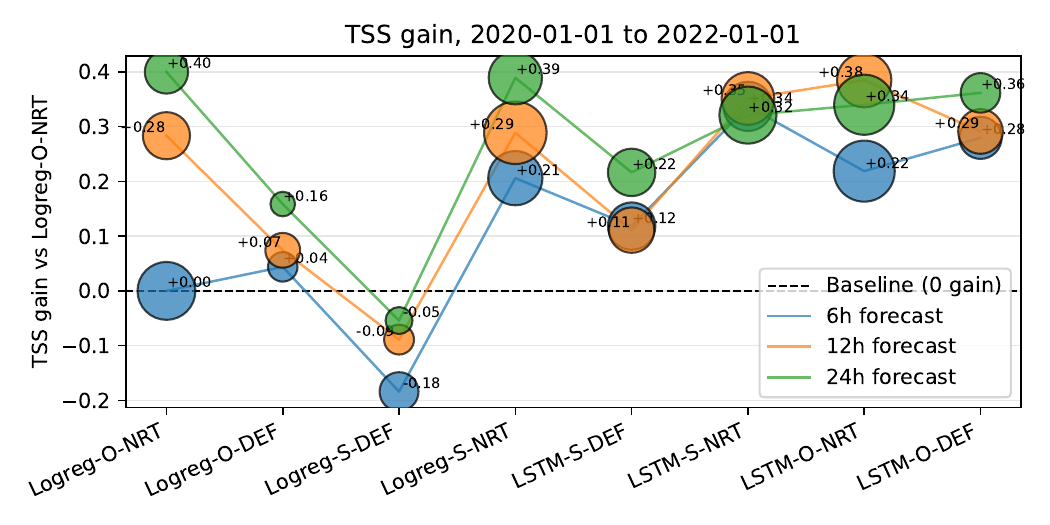}
        \caption{TSS, Solar Minimum}
    \end{subfigure}
    \begin{subfigure}[b]{0.48\textwidth}
        \centering
        \includegraphics[width=\textwidth]{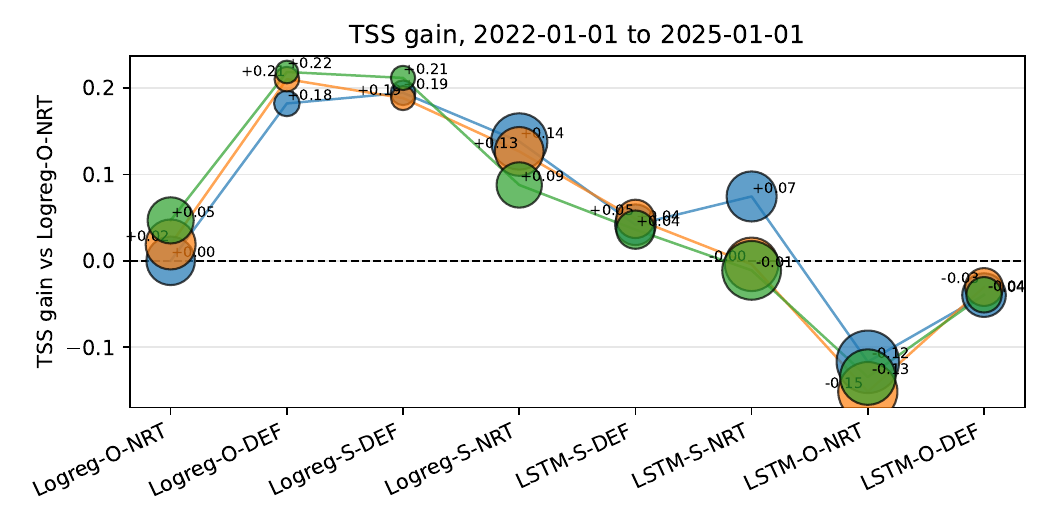}
        \caption{TSS, Evolving + Solar Maximum}
    \end{subfigure}
    \begin{subfigure}[b]{0.48\textwidth}
        \centering
        \includegraphics[width=\textwidth]{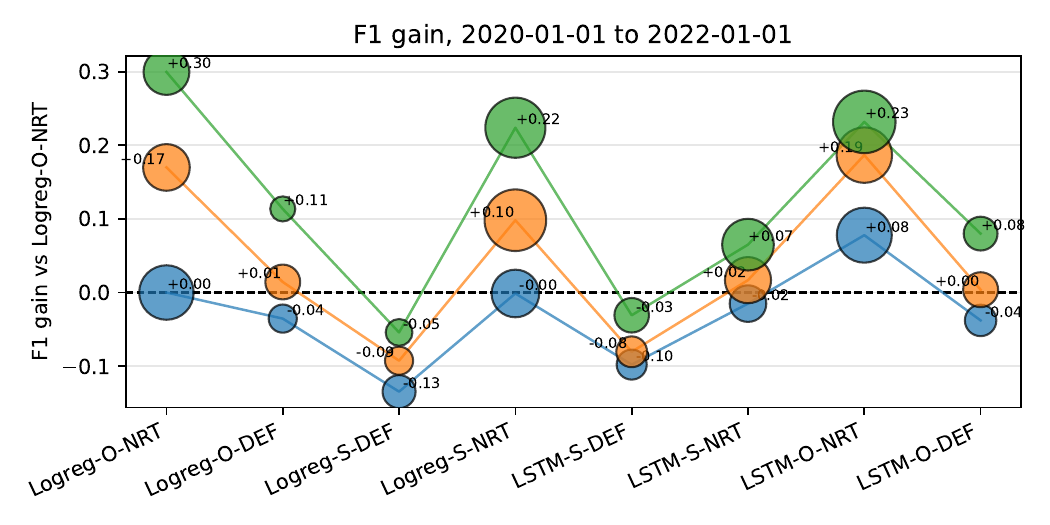}
        \caption{F1, Solar Minimum}
    \end{subfigure}
    \begin{subfigure}[b]{0.48\textwidth}
        \centering
        \includegraphics[width=\textwidth]{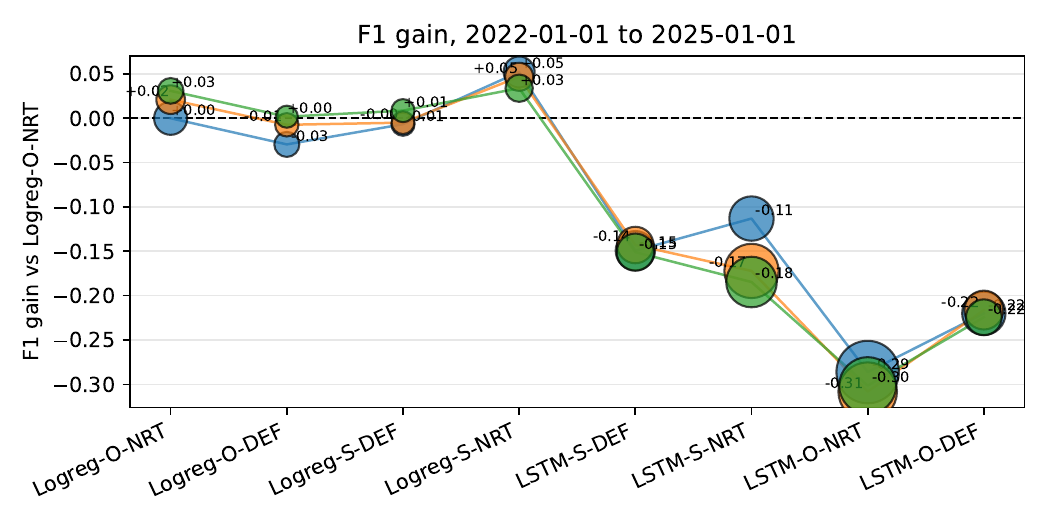}
        \caption{F1, Evolving + Solar Maximum}
    \end{subfigure}
    \caption{Average performance gain relative to the baseline Logreg-O-NRT model for Near-Real-Time SHARP predictors, evaluated during solar minimum (2020–2022) and Evolving + solar maximum (2022–2025) testing periods. The baseline model is the logistic regression model for 6-hour forecasting trained on \texttt{SWPC-FTP} flare lists and NRT-SHARPs (Logreg-O-NRT-6). Each marker represents a model configuration, with color indicating forecasting window length (6 h, 12 h, and 24 h). Marker size is proportional to the standard deviation across 30 bootstrap independent runs. Positive values indicate improvement over the baseline model (black dashed zero line).}
    \label{fig:metric_gain_nrtSHARPs}
\end{figure}

Figure~\ref{fig:metric_gain_nrtSHARPs} presents the average TSS and F1 score gains across eight model configurations. Because the NRT-SHARPs dataset is extremely imbalanced during 2023–2024, we compare performance between the solar minimum and the combined evolving plus solar maximum periods. 

During solar minimum, and holding other settings fixed, training on NRT-SHARPs generally improves performance for both logistic and LSTM models. In contrast, during the evolving plus solar maximum period, training on NRT-SHARPs tends to degrade performance overall, particularly in terms of TSS. {For F1 score, the logistic regression models exhibit similar performance across configurations. Training on NRT-SHARPs generally reduces F1 performance, with two exceptions: logistic regression models trained on the \texttt{Science-Quality} flare list, and the LSTM model trained on \texttt{Science-Quality} with a 6-hour forecasting window.} 
The results also demonstrate that models trained on NRT-SHARPs exhibit greater {performance} variability than those trained on DEF-SHARPs, {as reflected by the larger marker sizes.} In addition, the greater overlap of markers indicates that models trained on DEF-SHARPs exhibit more consistent performance across different forecasting-window lengths.

\subsection{Practical guidance for data-product selection}
\label{sec:practical_guidance}

{The comparisons in the previous Sections show that there is no universally optimal flare catalog or predictor product. The appropriate choice depends on the scientific or operational objective, whether active-region labels are required, whether real-time availability is necessary, and whether the priority is calibrated physical consistency or reproducibility of operational forecasting conditions. Table~\ref{tab:practical_guidance} summarizes the practical recommendations implied by the catalog comparisons, data-quality analyses, and forecasting experiments presented in this paper.}

\begin{table}[!htbp]
\centering
\caption{Recommendations for selecting flare-response catalogs and predictor products.}
\label{tab:practical_guidance}
\footnotesize
\renewcommand{\arraystretch}{1.15}
\begin{tabularx}{\textwidth}{
>{\raggedright\arraybackslash}p{0.16\textwidth}
>{\raggedright\arraybackslash}X
>{\raggedright\arraybackslash}X
>{\raggedright\arraybackslash}X
}
\toprule
\textbf{Data Product} &
\textbf{{Recommended Use Case(s)}} &
\textbf{{Required Processing and Documentation}} &
\textbf{{Key Limitation(s)}} \\
\midrule
SWPC-FTP GOES Flare List &
Operational forecasting; ML labels; comparisons with SWPC-style operational products &
Report satellite \& time range; correct or flag pre-GOES-R scaling when comparing intensities with science-quality data &
Operational/preliminary product; no precise flare location; pre-GOES-R magnitudes affected by SWPC scaling \\

\hline

NCEI Science-Quality GOES Flare Summary &
Retrospective scientific studies; calibrated flare magnitudes; GOES-R location-based analyses &
Use appropriate satellite periods; augment AR numbers if needed &
No NOAA AR numbers; incomplete AR assignment reduces total usable samples \\

\hline

Augmented Science-Quality Flare List &
Retrospective ML studies, when calibrated labels and AR matching are both desired &
{Report matching tolerances, method for assigning ARs, distance threshold, and AR coverage} &
Contains some invalid AR numbers; possible information loss in AR-based training \\

\hline

SSW Catalog &
Auxiliary event matching or location support &
Operational-style {product}; correct or account for pre-GOES-R SWPC scaling &
Has SWPC behavior before GOES-R; {should} not {be} used as a stand-alone benchmark \\

\hline

SunPy-HEK GOES Flare Records &
Exploratory event search only, after validation &
Cross-check event counts, magnitudes, and AR numbers against SWPC-FTP or NCEI before use &
Not recommended for ML training or benchmarking, especially for GOES-R-era data \\

\hline

DEF-SHARPs &
Retrospective ML benchmarks; scientific model comparison; stable predictor set &
Remove missing, incomplete, infinite, low-quality, and near-limb records &
Not available in real time; delayed relative to operational forecasting needs \\

\hline

NRT-SHARPs &
Real-time or operational forecasting &
Handle the reduced quality in data, smaller effective sample size, and a greater score variability&
Less stable than DEF-SHARPs; HARP identifiers are not consistent with DEF-SHARPs \\

\hline

HMI/AIA {AR} Image Predictors &
Image-based machine-learning studies when spatial structure is important &
Align channels; remove near-limb active regions; require sufficient pre-flare temporal coverage &
Near-limb images \& incomplete sequences can introduce systematic quality problems \\

\bottomrule
\end{tabularx}
\end{table}

{For machine-learning studies that require NOAA active-region identifiers to link flare labels to SHARP, HARP, or image predictors, the \texttt{SWPC-FTP} operational flare list remains the most direct response-label source because it includes active-region numbers natively. It is especially appropriate when the goal is to reproduce or evaluate operational forecasting behavior. However, for pre-GOES-R data, the SWPC scaling factor should be documented and corrected when comparisons are made with science-quality irradiances or when flare intensities are analyzed quantitatively.}

{For retrospective studies where calibrated flare magnitudes, homogeneous irradiance processing, or GOES-R flare locations are important, the NCEI science-quality flare products are preferable. However, these products do not directly provide NOAA active region numbers. Therefore, when science-quality labels are used in active-region-based flare prediction, active-region assignment must be performed explicitly and the resulting coverage should be reported. In our augmented \texttt{Science-Quality} list, valid active-region numbers can be assigned to a substantial fraction of flares, but not all of them; this information loss should be considered when interpreting machine-learning skill scores.}

{The \texttt{SSW} catalog should be used cautiously as an auxiliary catalog, especially for event matching or location information, but should not be treated as an independent science-quality target. Our comparison indicates that, before the GOES-R operational transition, \texttt{SSW} carries the SWPC scaling behavior and is therefore closer to an operational-style product. Similarly, the \texttt{SunPy-HEK} GOES flare list should not be used directly for training or benchmarking flare-forecasting models without independent validation, particularly for GOES-R-era data, because of event-count discrepancies and active-region-number inconsistencies relative to \texttt{SWPC-FTP}.}

{For SHARP predictor data, definitive SHARPs are recommended for retrospective machine learning benchmarks and scientific model comparisons because they incorporate improved calibration and processing and lead to more stable forecasting performance, especially for complex models such as LSTMs. Near-real-time SHARPs are appropriate when the goal is real-time or operational forecasting, but their reduced data quality, smaller effective sample size, inconsistent HARP identifiers relative to definitive SHARPs, and increased variability in model skill scores should be reported. For image-based predictors, active regions near the limb and image sequences with incomplete temporal coverage should be filtered or treated separately, because near-limb geometry and missing image frames can degrade model inputs.}

\section{Conclusion}
\label{sec:conclusion}

In this paper, we systematically review and compare the most commonly used data products for solar flare forecasting using machine learning methods. For the flare-response labels, we cross-evaluate the \texttt{Science-Quality} list from NCEI, the operational \texttt{SWPC-FTP} list, and the \texttt{SSW} catalog. We also construct an augmented Science-Quality flare list with assigned NOAA active-region numbers and develop a reproducible processing pipeline for generating this enhanced dataset. For predictor data, we describe the imaging products from AIA/SDO and HMI/SDO and analyze the quality issues that arise near the solar limb. We further examine completeness, missingness, and other quality concerns in the HARP and SHARP parameter datasets.

{Overall, the choice of data product should be matched to the forecasting objective rather than treated as a fixed universal preference. For operational active-region-based forecasting, SWPC-FTP labels and NRT-SHARPs most closely represent the real-time forecasting setting, but their operational limitations should be reported. For retrospective machine-learning benchmarks, DEF-SHARPs are preferable because they provide a more stable predictor product, while the choice between SWPC-FTP and science-quality flare labels depends on whether native active-region labels or calibrated flare magnitudes are more important. NCEI \texttt{Science-Quality} flare products are best suited to calibrated retrospective analyses, but active-region augmentation and the associated loss of unmatched events must be documented when they are used for active-region-level prediction. \texttt{SSW} and \texttt{SunPy-HEK} should be used with particular caution:  \texttt{SSW} behaves as an operational-style catalog before the GOES-R transition, and \texttt{SunPy-HEK} GOES records should be independently validated before being used in model training or evaluation.}

In addition, we compare the predictive performance of LSTM-based models and logistic-regression baselines across multiple combinations of flare lists (\texttt{Science-Quality} and \texttt{SWPC-FTP}) and SHARP data (DEF-SHARPs and NRT-SHARPs). Our analysis results in the following conclusions. 
\begin{enumerate}
    \item The \texttt{Science-Quality} flare list does not yield notable improvements in forecasting skill, particularly during solar minimum. This may be explained by information loss: only about 72\% of Science-Quality flares can be assigned a valid active region number and thus incorporated into model training.
    \item Near-real-time SHARPs, as compared with definitive SHARPs, introduce substantially greater variability in model skill scores, likely resulting from reduced data quality and a smaller effective training sample.
    \item The impact of data products is strongly solar-cycle dependent. Holding other factors fixed, the operational \texttt{SWPC-FTP} labels improve model performance during solar minimum but generally degrade performance during solar maximum. The NRT-SHARPs and DEF-SHARPs comparison exhibits a similar pattern.
    \item Although the more sophisticated LSTM models do not consistently outperform logistic regression, they produce more stable results across forecasting windows and data product combinations, indicating robustness to label and predictor variations.
\end{enumerate}

By clarifying the strengths and limitations of each product and quantifying their effects on predictive performance, we aim to help researchers select appropriate datasets and to improve the comparability and interpretability of data-driven flare-forecasting studies.

\section*{Data/Software Availability Statement}

All data are publicly available from SWPC (\url{https://www.ncei.noaa.gov/products/space-weather/partners/swpc-products-and-data}), JSOC (\url{http://jsoc.stanford.edu/HMI/HARPS.html}), and GOES (\url{https://www.ncei.noaa.gov/data/goes-space-environment-monitor/access/science/xrs/}). The code and results are available on GitHub \url{https://github.com/kehu001/CLEAR-Flare-Forecasting-Models}.

\section*{Acknowledgments}

The authors acknowledge NASA for funding the CLEAR Space Weather Center of Excellence (grant number 80NSSC24K0062).

\bibliographystyle{plainnat}
\bibliography{bibliography}

\appendix

\section{Quiet Period Analysis of Active Regions}
\begin{figure}[!htb]
    \centering
    \includegraphics[width=\textwidth]{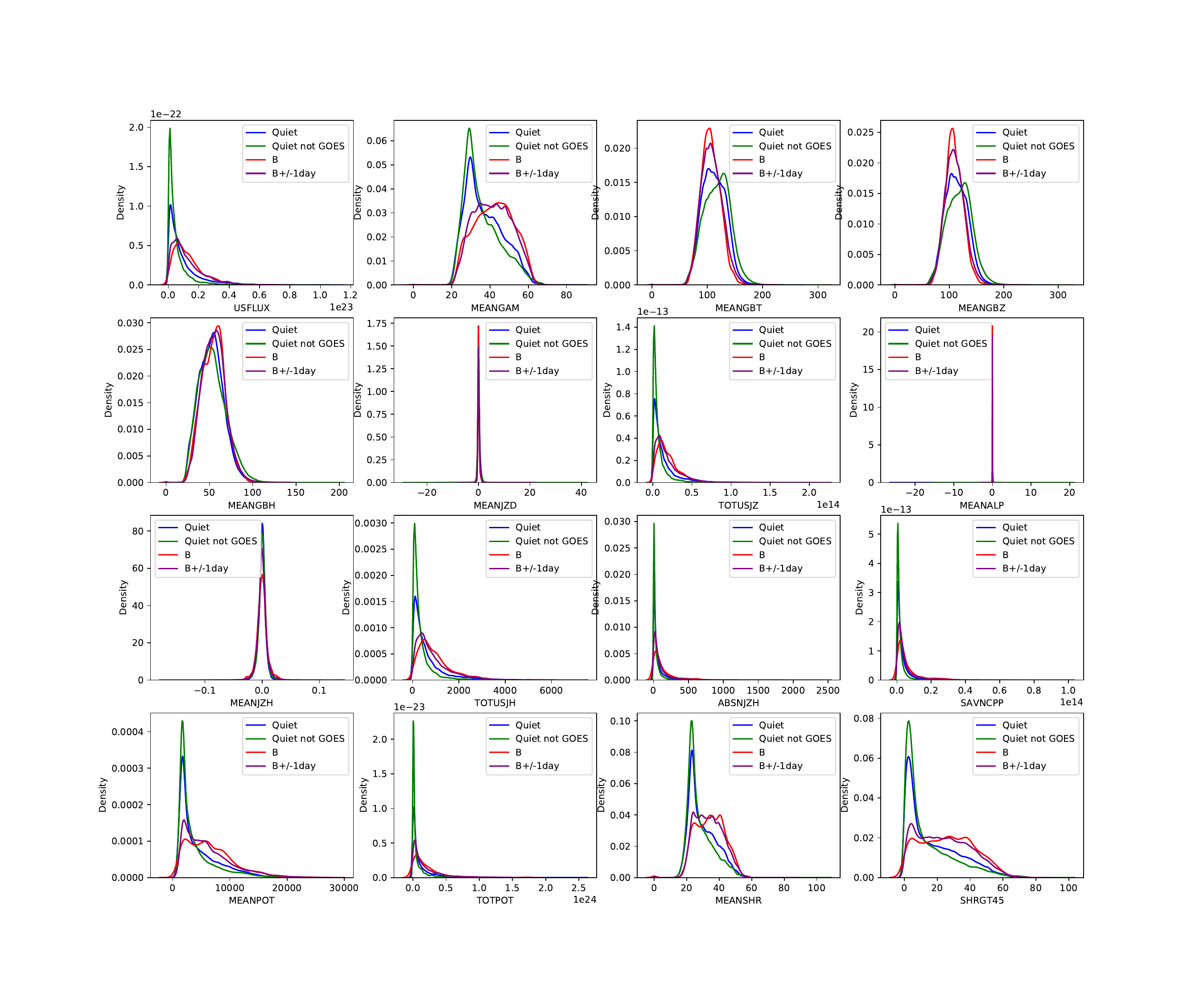}
    \caption{Marginal Distribution of SHARPs over different periods aggregated across Active Regions.}
    \label{fig:quiet_prop}
\end{figure}

In this section, we aim to determine whether there are distributional differences between active regions during ongoing flares and during a ``quiet period''. As an initial approach, we define the "quiet period" of an active region as any period 24 hours removed from flare activity (i.e., more than 24 hours before the flare start time and after the flare ends). We measure these distributional differences by examining the marginal distribution of all SHARPs, depending on whether it's a quiet period or a flaring period (defined as any time that is not a quiet period).  After reviewing all flares from 2010 to 2024 and categorizing the periods of all active regions during this interval as quiet or flaring, we analyzed the marginal distribution of SHARPs during this period. These are visualized here in Figure \ref{fig:quiet_prop}

The quiet (blue) line depicts the SHARP parameters during the quiet period of an AR. The green line depicts the SHARP parameters of active regions that are available in the combined SHARPs data but do not have any flares associated with them in the GOES flare list. For the active regions present in the GOES flare list, we also visualized the distribution of the SHARPs during B-flares to investigate if B-flares could serve as a ```surrogate'' for quiet time. The red curve depicts the exact time interval of a B-flare as described by the GOES flare list, and the purple line expands the interval by 24 hours before and after the start and end times, respectively. This plot shows that the distribution of a few SHARPs differs between quiet periods and periods of activity. In particular, the distributions of MEANGAM, MEANSHR, and SHGRT45 suggest that active regions exhibit different behavior across time periods. To better generalize the behavior of active regions, it may be beneficial to incorporate quiet-time data from active regions as well.

\begin{figure}[!htb]
    \centering
    \includegraphics[width=\textwidth]{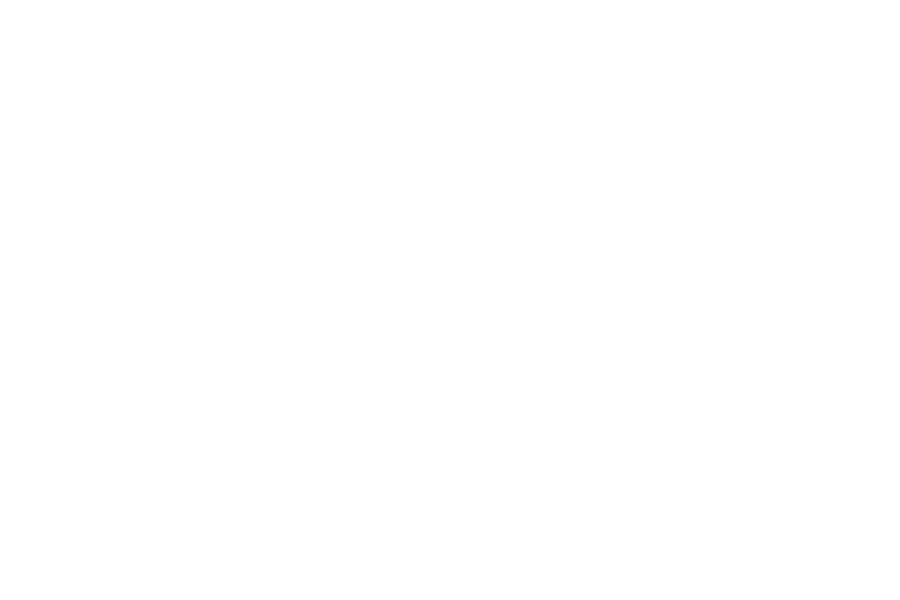}
    \caption{Marginal Distribution of SHARPs over different periods aggregated across Active Regions.}
    \label{fig:96-24flares}
\end{figure}

Additionally, Figure \ref{fig:96-24flares} depicts the class of all flares in the GOES flare list starting from 1996, which encapsulates two solar cycles. At the peak of the solar cycle, a percentage of B-flares is missing. This indicates that the flare list may not capture all B-flares, as they could be obscured by stronger flares that occur more frequently during the solar cycle. As such, quiet-time imaging data from these solar-cycle peaks can also serve as a representation of the B-flares during this period. 
 
Additionally, the two dashed lines represent the potential shift that is necessary for the GOES flare labels. As discussed in Section \ref{sec:Flare_Response}, there was an erroneous scaling factor being applied to flares measured prior to the GOES-16 satellite. If we adjust for this scaling, all the flares from before 2019 will be shifted up the length from the ``black" line'' to the red line. In this scenario, we will have systematic missingness of low-intensity B-flares in our data; therefore, having quiet data during this period can help mitigate potential negative effects.

\newpage
\section{Comparison of Numerical Results}
\label{appendix:comparisonresults}

\begin{table}[!ht]
\centering
\scriptsize  
\setlength{\tabcolsep}{3.5pt}
\caption{Summary of LSTM model performance for 24-hr forecasting using definitive SHARP parameters with \texttt{Science-Quality} flares (S-DEF) and \texttt{SWPC-FTP} flares (O-DEF) across three independent testing periods. Each cell reports the mean and range (minimum–maximum) of bootstrap ensemble results. The rows correspond to varying levels of solar activity: ``Min'' (\mbox{2020--2021}), ``Evolving'' (2022), ``Max'' (\mbox{2023--2024}), and ``All'' (2020-2024).}
\label{tab:lstm_24_sci_opr_defSHARPS_1}
\begin{tabular}{lccccccc}
\toprule
\textbf{Period} & \textbf{Model} & \textbf{TSS} & \textbf{HSS} & \textbf{POD} & \textbf{F1} & \textbf{FAR} & \textbf{ACC} \\
\midrule
\multirow{2}{*}{\textbf{Min}}
& S-DEF & 0.42 [0.16, 0.59] & 0.09 [0.03, 0.13] & 0.75 [0.45, 0.91] & 0.15 [0.10, 0.19] & 0.32 [0.28, 0.39] & 0.68 [0.62, 0.72] \\
& O-DEF & \textbf{0.57 [0.40, 0.69]} & \textbf{0.19 [0.14, 0.24]} & \textbf{0.82 [0.64, 0.96]} & \textbf{0.26 [0.21, 0.30]} & \textbf{0.25 [0.21, 0.30]} & \textbf{0.75 [0.71, 0.79]} \\
\midrule
\multirow{2}{*}{\textbf{Evolving}}
& S-DEF & 0.59 [0.50, 0.68] & 0.60 [0.51, 0.69] & \textbf{0.71 [0.62, 0.78]} & 0.73 [0.66, 0.79] & 0.12 [0.07, 0.20] & 0.82 [0.78, 0.87] \\
& O-DEF & \textbf{0.58 [0.53, 0.66]} & \textbf{0.61 [0.56, 0.69]} & 0.64 [0.58, 0.72] & \textbf{0.73 [0.70, 0.79]} & \textbf{0.06 [0.03, 0.10]} & \textbf{0.82 [0.80, 0.86]} \\
\midrule
\multirow{2}{*}{\textbf{Max}}
& S-DEF & \textbf{0.44 [0.39, 0.50]} & \textbf{0.09 [0.07, 0.11]} & \textbf{0.46 [0.40, 0.52]} & \textbf{0.63 [0.57, 0.69]} & \textbf{0.01 [0.00, 0.04]} & \textbf{0.49 [0.43, 0.55]} \\
& O-DEF & 0.38 [0.31, 0.44] & 0.05 [0.04, 0.07] & 0.38 [0.31, 0.44] & 0.55 [0.48, 0.61] & 0.00 [0.00, 0.00] & 0.41 [0.34, 0.47] \\
\midrule
\multirow{2}{*}{\textbf{All}}
& S-DEF & \textbf{0.57 [0.49, 0.62]} & \textbf{0.57 [0.50, 0.63]} & \textbf{0.63 [0.56, 0.71]} & \textbf{0.74 [0.69, 0.79]} & 0.06 [0.04, 0.10] & \textbf{0.79 [0.76, 0.82]} \\
& O-DEF & 0.51 [0.46, 0.58] & 0.52 [0.46, 0.58] & 0.55 [0.49, 0.61] & 0.69 [0.65, 0.74] & \textbf{0.03 [0.01, 0.05]} & 0.76 [0.73, 0.79] \\
\bottomrule
\end{tabular}
\end{table}

\begin{table}[!ht]
\centering
\scriptsize  
\setlength{\tabcolsep}{3.5pt}
\caption{Summary of LSTM model performance for 12-hr forecasting using definitive SHARP parameters with \texttt{Science-Quality} flares (S-DEF) and \texttt{SWPC-FTP} flares (O-DEF) across three independent testing periods. Each cell reports the mean and range (minimum–maximum) of bootstrap ensemble results. The rows correspond to varying levels of solar activity: ``Min'' (\mbox{2020--2021}), ``Evolving'' (2022), ``Max'' (\mbox{2023--2024}), and ``All'' (2020-2024).}
\label{tab:lstm_12_sci_opr_defSHARPS}
\begin{tabular}{lccccccc}
\toprule
\textbf{Period} & \textbf{Model} & \textbf{TSS} & \textbf{HSS} & \textbf{POD} & \textbf{F1} & \textbf{FAR} & \textbf{ACC} \\
\midrule
\multirow{2}{*}{2020--2022}
& S-DEF & 0.32 [0.07, 0.45] & 0.05 [0.01, 0.07] & 0.65 [0.38, 0.82] & 0.10 [0.06, 0.12] & 0.33 [0.30, 0.37] & 0.67 [0.63, 0.70] \\
& O-DEF & \textbf{0.50 [0.34, 0.64]} & \textbf{0.12 [0.08, 0.18]} & \textbf{0.77 [0.61, 0.92]} & \textbf{0.18 [0.15, 0.23]} & \textbf{0.27 [0.23, 0.31]} & \textbf{0.73 [0.70, 0.78]} \\
\midrule
\multirow{2}{*}{2022--2023}
& S-DEF & 0.59 [0.53, 0.70] & 0.58 [0.52, 0.67] & \textbf{0.72 [0.63, 0.83]} & 0.70 [0.66, 0.77] & 0.13 [0.07, 0.17] & 0.83 [0.80, 0.87] \\
& O-DEF & \textbf{0.59 [0.52, 0.63]} & \textbf{0.63 [0.54, 0.68]} & 0.65 [0.61, 0.71] & \textbf{0.74 [0.68, 0.77]} & \textbf{0.06 [0.03, 0.11]} & \textbf{0.84 [0.80, 0.87]} \\
\midrule
\multirow{2}{*}{2023--2025}
& S-DEF & \textbf{0.44 [0.38, 0.50]} & \textbf{0.10 [0.08, 0.12]} & \textbf{0.45 [0.39, 0.52]} & \textbf{0.62 [0.56, 0.68]} & \textbf{0.01 [0.00, 0.02]} & \textbf{0.48 [0.43, 0.55]} \\
& O-DEF & 0.37 [0.30, 0.45] & 0.06 [0.04, 0.08] & 0.37 [0.31, 0.45] & 0.54 [0.47, 0.62] & 0.00 [0.00, 0.02] & 0.41 [0.34, 0.48] \\
\midrule
\multirow{2}{*}{2020--2025}
& S-DEF & \textbf{0.58 [0.50, 0.66]} & \textbf{0.60 [0.52, 0.67]} & \textbf{0.65 [0.56, 0.73]} & \textbf{0.75 [0.68, 0.80]} & 0.06 [0.04, 0.08] & \textbf{0.81 [0.77, 0.84]} \\
& O-DEF & 0.53 [0.47, 0.58] & 0.55 [0.49, 0.60] & 0.57 [0.50, 0.63] & 0.70 [0.65, 0.74] & \textbf{0.04 [0.02, 0.07]} & 0.79 [0.76, 0.81] \\
\bottomrule

\end{tabular}
\end{table}

\begin{table}[!ht]
\centering
\scriptsize  
\setlength{\tabcolsep}{3.5pt}
\caption{Summary of LSTM model performance for 6-hr forecasting using definitive SHARP parameters with \texttt{Science-Quality} flares (S-DEF) and \texttt{SWPC-FTP} flares (O-DEF) across three independent testing periods. Each cell reports the mean and range (minimum–maximum) of bootstrap ensemble results. The rows correspond to varying levels of solar activity: ``Min'' (\mbox{2020--2021}), ``Evolving'' (2022), ``Max'' (\mbox{2023--2024}), and ``All'' (2020-2024).}
\label{tab:lstm_6_sci_opr_defSHARPS}
\begin{tabular}{lccccccc}
\toprule
\textbf{Period} & \textbf{Model} & \textbf{TSS} & \textbf{HSS} & \textbf{POD} & \textbf{F1} & \textbf{FAR} & \textbf{ACC} \\
\midrule
\multirow{2}{*}{2020--2022}
& S-DEF & 0.33 [0.03, 0.48] & 0.04 [0.00, 0.06] & 0.66 [0.37, 0.85] & 0.08 [0.05, 0.10] & 0.33 [0.27, 0.37] & 0.67 [0.63, 0.72] \\
& O-DEF & \textbf{0.49 [0.30, 0.60]} & \textbf{0.09 [0.05, 0.13]} & \textbf{0.77 [0.60, 0.93]} & \textbf{0.14 [0.11, 0.17]} & \textbf{0.29 [0.24, 0.34]} & \textbf{0.71 [0.66, 0.77]} \\
\midrule
\multirow{2}{*}{2022--2023}
& S-DEF & 0.57 [0.52, 0.63] & 0.55 [0.49, 0.61] & \textbf{0.73 [0.65, 0.79]} & 0.67 [0.63, 0.72] & 0.15 [0.10, 0.19] & 0.82 [0.79, 0.85] \\
& O-DEF & \textbf{0.60 [0.52, 0.66]} & \textbf{0.62 [0.55, 0.69]} & 0.69 [0.60, 0.79] & \textbf{0.72 [0.67, 0.77]} & \textbf{0.09 [0.05, 0.16]} & \textbf{0.84 [0.82, 0.88]} \\
\midrule
\multirow{2}{*}{2023--2025}
& S-DEF & \textbf{0.43 [0.34, 0.47]} & \textbf{0.11 [0.08, 0.13]} & \textbf{0.43 [0.35, 0.48]} & \textbf{0.61 [0.52, 0.65]} & \textbf{0.01 [0.00, 0.02]} & \textbf{0.48 [0.40, 0.52]} \\
& O-DEF & 0.36 [0.25, 0.48] & 0.07 [0.04, 0.10] & 0.36 [0.26, 0.48] & 0.53 [0.41, 0.65] & 0.00 [0.00, 0.04] & 0.40 [0.31, 0.51] \\
\midrule
\multirow{2}{*}{2020--2025}
& S-DEF & \textbf{0.58 [0.48, 0.65]} & \textbf{0.60 [0.51, 0.67]} & \textbf{0.66 [0.56, 0.72]} & \textbf{0.74 [0.67, 0.79]} & 0.08 [0.05, 0.11] & \textbf{0.82 [0.78, 0.84]} \\
& O-DEF & 0.53 [0.49, 0.64] & 0.56 [0.51, 0.65] & 0.60 [0.52, 0.74] & 0.71 [0.66, 0.79] & \textbf{0.06 [0.03, 0.10]} & 0.80 [0.78, 0.84] \\
\bottomrule

\end{tabular}
\end{table}

\begin{table}[!ht]
\centering
\scriptsize  
\setlength{\tabcolsep}{3.5pt}
\caption{Summary of Logistic model performance for 24-hr forecasting using definitive SHARP parameters with \texttt{Science-Quality} flares (S-DEF) and \texttt{SWPC-FTP} flares (O-DEF) across three independent testing periods. Each cell reports the mean and range (minimum–maximum) of bootstrap ensemble results. The rows correspond to varying levels of solar activity: ``Min'' (\mbox{2020--2021}), ``Evolving'' (2022), ``Max'' (\mbox{2023--2024}), and ``All'' (2020-2024).}
\label{tab:lstm_24_sci_opr_defSHARPS_2}
\begin{tabular}{lccccccc}
\toprule
\textbf{Period} & \textbf{Model} & \textbf{TSS} & \textbf{HSS} & \textbf{POD} & \textbf{F1} & \textbf{FAR} & \textbf{ACC} \\
\midrule
\multirow{2}{*}{Min} 
& Logistic-S-DEF & 0.15 [0.11, 0.18] & 0.07 [0.06, 0.09] & 0.27 [0.23, 0.30] & 0.13 [0.11, 0.14] & 0.12 [0.11, 0.13] & 0.86 [0.84, 0.87] \\
& Logistic-O-DEF & 0.37 [0.35, 0.38] & 0.24 [0.23, 0.25] & 0.46 [0.43, 0.48] & 0.29 [0.29, 0.30] & 0.09 [0.09, 0.10] & 0.88 [0.88, 0.89] \\
\midrule
\multirow{2}{*}{Evolving} 
& Logistic-S-DEF & 0.62 [0.61, 0.64] & 0.65 [0.64, 0.67] & 0.69 [0.69, 0.71] & 0.75 [0.75, 0.77] & 0.07 [0.07, 0.07] & 0.85 [0.85, 0.86] \\
& Logistic-O-DEF & 0.68 [0.67, 0.69] & 0.71 [0.70, 0.71] & 0.73 [0.72, 0.74] & 0.81 [0.80, 0.81] & 0.05 [0.05, 0.06] & 0.87 [0.86, 0.87] \\
\midrule
\multirow{2}{*}{Max} 
& Logistic-S-DEF & 0.69 [0.67, 0.70] & 0.27 [0.26, 0.28] & 0.79 [0.78, 0.80] & 0.88 [0.87, 0.89] & 0.10 [0.09, 0.13] & 0.80 [0.79, 0.81] \\
& Logistic-O-DEF & 0.74 [0.72, 0.76] & 0.21 [0.21, 0.22] & 0.76 [0.76, 0.77] & 0.87 [0.86, 0.87] & 0.02 [0.01, 0.05] & 0.77 [0.77, 0.78] \\
\midrule
\multirow{2}{*}{All} 
& Logistic-S-DEF & 0.66 [0.65, 0.67] & 0.66 [0.65, 0.67] & 0.76 [0.75, 0.77] & 0.81 [0.81, 0.82] & 0.10 [0.10, 0.11] & 0.83 [0.83, 0.84] \\
& Logistic-O-DEF & 0.67 [0.66, 0.67] & 0.67 [0.66, 0.67] & 0.75 [0.74, 0.75] & 0.82 [0.81, 0.82] & 0.08 [0.07, 0.08] & 0.83 [0.83, 0.84] \\
\bottomrule
\end{tabular}
\end{table}

\begin{table}[!ht]
\centering
\scriptsize  
\setlength{\tabcolsep}{3.5pt}
\caption{Summary of Logistic model performance for 12-hr forecasting using definitive SHARP parameters with \texttt{Science-Quality} flares (S-DEF) and \texttt{SWPC-FTP} flares (O-DEF) across three independent testing periods. Each cell reports the mean and range (minimum–maximum) of bootstrap ensemble results. The rows correspond to varying levels of solar activity: ``Min'' (\mbox{2020--2021}), ``Evolving'' (2022), ``Max'' (\mbox{2023--2024}), and ``All'' (2020-2024).}
\label{tab:lstm_24_sci_opr_defSHARPS_3}
\begin{tabular}{lccccccc}
\toprule
\textbf{Period} & \textbf{Model} & \textbf{TSS} & \textbf{HSS} & \textbf{POD} & \textbf{F1} & \textbf{FAR} & \textbf{ACC} \\
\midrule
\multirow{2}{*}{\textbf{Min}}
& S-DEF & 0.12 [0.05, 0.17] & 0.04 [0.02, 0.06] & 0.25 [0.18, 0.30] & 0.09 [0.06, 0.11] & 0.13 [0.12, 0.13] & 0.86 [0.85, 0.86] \\
& O-DEF & \textbf{0.28 [0.22, 0.38]} & \textbf{0.14 [0.11, 0.19]} & \textbf{0.39 [0.32, 0.49]} & \textbf{0.19 [0.17, 0.24]} & \textbf{0.11 [0.10, 0.11]} & \textbf{0.87 [0.87, 0.88]} \\
\midrule
\multirow{2}{*}{\textbf{Evolving}}
& S-DEF & 0.58 [0.56, 0.61] & 0.60 [0.58, 0.64] & 0.65 [0.63, 0.68] & 0.71 [0.69, 0.73] & 0.08 [0.07, 0.09] & 0.85 [0.84, 0.86] \\
& O-DEF & \textbf{0.64 [0.62, 0.66]} & \textbf{0.68 [0.66, 0.70]} & \textbf{0.69 [0.66, 0.71]} & \textbf{0.77 [0.76, 0.79]} & \textbf{0.04 [0.04, 0.05]} & \textbf{0.87 [0.86, 0.87]} \\
\midrule
\multirow{2}{*}{\textbf{Max}}
& S-DEF & 0.66 [0.65, 0.67] & \textbf{0.28 [0.28, 0.29]} & \textbf{0.78 [0.78, 0.79]} & \textbf{0.87 [0.87, 0.88]} & 0.12 [0.11, 0.13] & \textbf{0.79 [0.78, 0.80]} \\
& O-DEF & \textbf{0.71 [0.67, 0.73]} & 0.23 [0.22, 0.24] & 0.76 [0.75, 0.77] & 0.86 [0.86, 0.87] & \textbf{0.05 [0.03, 0.08]} & 0.77 [0.76, 0.78] \\
\midrule
\multirow{2}{*}{\textbf{All}}
& S-DEF & 0.64 [0.63, 0.65] & 0.64 [0.64, 0.65] & \textbf{0.75 [0.74, 0.76]} & 0.79 [0.79, 0.80] & 0.11 [0.10, 0.12] & 0.83 [0.82, 0.83] \\
& O-DEF & \textbf{0.65 [0.64, 0.65]} & \textbf{0.66 [0.65, 0.66]} & 0.73 [0.72, 0.74] & \textbf{0.80 [0.79, 0.80]} & \textbf{0.09 [0.08, 0.09]} & \textbf{0.83 [0.83, 0.84]} \\
\bottomrule
\end{tabular}
\end{table}

\begin{table}[!ht]
\centering
\scriptsize  
\setlength{\tabcolsep}{3.5pt}
\caption{Summary of Logistic model performance for 6-hr forecasting using definitive SHARP parameters with \texttt{Science-Quality} flares (S-DEF) and \texttt{SWPC-FTP} flares (O-DEF) across three independent testing periods. Each cell reports the mean and range (minimum–maximum) of bootstrap ensemble results. The rows correspond to varying levels of solar activity: ``Min'' (\mbox{2020--2021}), ``Evolving'' (2022), ``Max'' (\mbox{2023--2024}), and ``All'' (2020-2024).}
\label{tab:lstm_24_sci_opr_defSHARPS_4}
\begin{tabular}{lccccccc}
\toprule
\textbf{Period} & \textbf{Model} & \textbf{TSS} & \textbf{HSS} & \textbf{POD} & \textbf{F1} & \textbf{FAR} & \textbf{ACC} \\
\midrule
\multirow{2}{*}{\textbf{Min}}
& S-DEF & 0.02 [-0.10, 0.20] & 0.01 [-0.03, 0.06] & 0.16 [0.04, 0.33] & 0.05 [0.01, 0.09] & 0.14 [0.13, 0.15] & 0.84 [0.83, 0.86] \\
& O-DEF & \textbf{0.25 [0.13, 0.27]} & \textbf{0.10 [0.06, 0.11]} & \textbf{0.37 [0.24, 0.38]} & \textbf{0.14 [0.10, 0.16]} & \textbf{0.12 [0.10, 0.13]} & \textbf{0.87 [0.85, 0.88]} \\
\midrule
\multirow{2}{*}{\textbf{Evolving}}
& S-DEF & 0.52 [0.48, 0.57] & 0.56 [0.53, 0.61] & 0.59 [0.54, 0.62] & 0.66 [0.63, 0.70] & 0.07 [0.06, 0.08] & 0.84 [0.84, 0.86] \\
& O-DEF & \textbf{0.55 [0.52, 0.59]} & \textbf{0.60 [0.57, 0.64]} & 0.59 [0.55, 0.64] & \textbf{0.70 [0.67, 0.73]} & \textbf{0.04 [0.03, 0.05]} & \textbf{0.85 [0.84, 0.86]} \\
\midrule
\multirow{2}{*}{\textbf{Max}}
& S-DEF & \textbf{0.67 [0.63, 0.71]} & \textbf{0.33 [0.31, 0.34]} & \textbf{0.79 [0.78, 0.80]} & \textbf{0.88 [0.87, 0.88]} & 0.12 [0.07, 0.15] & \textbf{0.80 [0.79, 0.80]} \\
& O-DEF & 0.68 [0.67, 0.72] & 0.24 [0.23, 0.26] & 0.74 [0.72, 0.76] & 0.85 [0.84, 0.86] & \textbf{0.06 [0.03, 0.09]} & 0.75 [0.73, 0.77] \\
\midrule
\multirow{2}{*}{\textbf{All}}
& S-DEF & \textbf{0.63 [0.62, 0.64]} & \textbf{0.64 [0.63, 0.65]} & \textbf{0.75 [0.73, 0.76]} & \textbf{0.78 [0.77, 0.78]} & 0.12 [0.11, 0.13] & \textbf{0.83 [0.82, 0.83]} \\
& O-DEF & 0.61 [0.59, 0.63] & 0.63 [0.61, 0.64] & 0.70 [0.68, 0.72] & 0.76 [0.75, 0.77] & \textbf{0.09 [0.08, 0.10]} & 0.82 [0.82, 0.83] \\
\bottomrule
\end{tabular}
\end{table}

\end{document}